\documentclass[12pt]{article}

\usepackage{setspace}

\usepackage{amsmath}
\usepackage{graphicx}
\usepackage{natbib}
\usepackage{url} 
\usepackage{latexsym}
\usepackage{amsfonts}
\usepackage{amssymb}
\usepackage{psfrag}
\usepackage{graphicx}
\usepackage{calc, cancel}
\usepackage{graphics, graphicx, caption, subcaption}
\usepackage{latexsym, pgfplots, tikz, url,verbatim,multicol,enumerate,bm,listings,color}
\usepackage{multirow}
\usepackage[justification=centering]{caption}

\usepackage{wrapfig, sidecap,enumitem}
\usepackage{epsfig}
\usepackage{longtable}

\newcommand{\normal}{\mathop{\rm N}}

\newcommand{\IG}{\mathop{\rm IG}}
\newcommand{\GIG}{\mathop{\rm GIG}}

\newcommand{\sumn}{\sum_{i = 1}^{N}}

\newcommand{\betav}{\mbox{\boldmath{$\beta$}}}

\newcommand{\muv}{\mbox{\boldmath{$\mu$}}}

\newcommand{\Sigmav}{\mbox{\boldmath{$\Sigma$}}}

\newcommand{\Lambdav}{\mbox{\boldmath{$\Lambda$}}}

\newcommand{\thetav}{\mbox{\boldmath{$\theta$}}}
\newcommand{\piv}{\mbox{\boldmath{$\pi$}}}

\def\half{\frac{1}{2}}
\def\E{\mathbb E}
\def\Id{\mathbb I} 

\def\Xv{\mathbf X}

\def\Tv{\mathbf T}

\def\uv{\mathbf u}
\def\Vv{\mathbf V}

\def\av{\mathbf a}

\def\cv{\mathbf c}
\def\xv{\mathbf x}

\def\zv{\mathbf z}
\def\tv{\mathbf t}

\def\1v{\mathbf 1}
\def\0v{\mathbf 0}

\begin{document}
	\doublespacing
	
	\title{Infinite mixtures of multivariate normal-inverse Gaussian distributions for clustering of skewed data}
	
	\author{Yuan Fang\footnote{Department of Mathematical Sciences, Binghamton University, State University of New York, 4400 Vestal Parkway East, Binghamton, NY, USA 13902. e: yfang8@binghamton.edu} \and Dimitris Karlis \footnote{Department of Statistics, University of Waterloo, thens University of Economics and Business, Athens, Greece. e: karlis@aueb.gr} \and Sanjeena Subedi \footnote{Department of Mathematical Sciences, Binghamton University, State University of New York, 4400 Vestal Parkway East, Binghamton, NY, USA 13902. e: sdang@binghamton.edu}}

	\maketitle

\begin{abstract}
	Mixtures of multivariate normal inverse Gaussian (MNIG)
	distributions can be used to cluster data that exhibit features
	such as skewness and heavy tails. However, for cluster analysis,
	using a traditional finite mixture model framework, either the
	number of components needs to be known {\it a-priori} or needs to
	be estimated {\it a-posteriori} using some model selection
	criterion after deriving results for a range of possible number of
	components. However, different model selection criteria can
	sometimes result in different number of components yielding
	uncertainty. Here, an infinite mixture model framework, also known
	as Dirichlet process mixture model, is proposed for the mixtures
	of MNIG distributions. This Dirichlet process mixture model
	approach allows the number of components to grow or decay freely
	from 1 to $\infty$ (in practice from 1 to $N$) and the number of
	components is inferred along with the parameter estimates in a
	Bayesian framework thus alleviating the need for model selection
	criteria. We provide real data applications with benchmark
	datasets as well as a small simulation experiment to compare with
	other existing models. The proposed method provides competitive
	clustering results to other clustering approaches for both
	simulation and real data and parameter recovery are illustrated
	using simulation studies.
\end{abstract}


\section{Introduction}
Finite mixture models, which assume that the population consists
of  a finite number of subpopulations, each represented by a known
distribution. Such models are commonly used for model-based
clustering purposes. In the recent literature, skewed mixture
models, which are based on non-symmetric marginal distributions
have been widely used and well studied. They can model both skewed
and symmetric components and are more robust to outliers. Some
examples include mixtures of skew-normal distributions
\citep{Lin2007norm}, mixtures of skew-$t$ distributions
\citep{Lin2007t,Pyne2009,Lin2010,Sylvia2010,vrbik2012,murray2014},
mixtures of generalized hyperbolic distributions
\citep{MGHD,wei2019mixtures}, mixtures of variance-gamma
distributions \citep{mcnicholas2017mixture}, and mixtures of
multivariate normal inverse Gaussian distributions
\citep{Karlis2009,Subedi2014,o2016clustering}.\par

Mixtures of multivariate normal inverse Gaussian distributions
(hereafter MNIG)  were  first proposed by \cite{Karlis2009} and
parameter estimation was done in an expectation-maximization (EM)
framework. \cite{Subedi2014} implemented an alternative
variational Bayes framework for parameter estimation for these
MNIG mixtures. Some well-known problems with EM algorithm for
mixture model parameter estimation can include slow convergence
and unreliable results arising from an unpleasant likelihood
surface, discussed by \cite{titterington1985}. A variational
inference tends to reach convergence faster and can be easily
scaled for larger datasets, yet is still an approximation to the
true posterior with not well understood statistical properties,
and does not provide exact coverage\citep{blei2017}.
\cite{fruhwirth2006finite} provides a detailed overview of the
Bayesian framework for modeling finite mixtures of distributions.
However, it is typically the case that the true number of groups
is unknown. Like other approaches to model-based clustering
through finite mixture models, parametric Bayesian approach cannot
easily overcome the problem to infer the number of components. It
either requires a pre-specified number of components, or needs the
employment of an information criterion for selecting the number of
components {\it a-posteriori}. Also one can use a Reversible
Jump MCMC to infer on the number of componenets
(see \cite{dellaportas2006multivariate}). While Bayesian
information criteria \citep[BIC;][]{schwarz1978} still remains the
most popular model selection criteria, the search of highly
effective model selection criteria, especially when dealing with
skewed data, still remains an open question.\par

Dirichlet process mixture models, also known as the  infinite
mixture models \citep{ferguson1973,antoniak1974}, is a
nonparametric Bayesian approach that allows the number of
components to vary in the model as a free parameter from $1$ to
$\infty$ (in practice from 1 to $N$, $N$ being the sample
size) by putting a Dirichlet process prior on the mixing
proportions \citep{west1992,BDA3,muller2013}. Hence, the number of
component is inferred simultaneously during parameter estimation.
The Dirichlet process mixture models have been applied in a wide
variety of applied problems
\citep{lartillot2004,huelsenbeck2007,onogi2011,hakguder2017}.
Early research on developing the Dirichlet process mixture models
goes back to three decades
\citep{west1992,escobar1995,maceachern1998}. \cite{neal2000} and
\cite{ishwaran2001} proposed two Markov chain Monte Carlo (MCMC)
frameworks for sampling from the posterior distributions of the
Dirichlet process mixture models.  In the context of model-based
clustering, \cite{rasmussen2000} proposed an infinite mixture of
Gaussian distribution; \cite{sivaganesan2002} adopted the Gibbs
sampling framework and implement the Dirichlet process mixture of
Gaussian distributions to cluster gene expression profiles. More
recent work includes \cite{wei2012} for an infinite Student's
t-mixture; \cite{sun2017} for a Dirichlet process mixture of
multivariate Gaussian distributions for clustering longitudinal
gene expression data; and \cite{hejblum2019sequential}, which
develops a sequential Dirichlet process mixtures of multivariate
skew $t-$distributions for clustering of flow cytometry data.
\cite{muller2013} provides a detailed review on the methodology,
application as well as generalizations of Dirichlet process
mixture models.

In this paper, we propose the infinite mixture framework for MNIG
distributions, and then illustrate clustering performance and
parameter estimation via a Gibbs sampling framework. The paper is
structured as follows: Section \ref{model} describes the finite
mixture of multivariate normal-inverse Gaussian distributions and
then extension to Dirichlet process mixture of MNIG distributions.
Section \ref{post} provides the posterior distributions for the
parameters for each hierarchical layer of the Dirichlet process
mixture model of MNIG distributions, followed by a Gibbs sampling
approach to posterior estimation with the discussion on
convergence diagnostics and label-switching issue in Section
\ref{gibbs}. In Section \ref{results}, competitive clustering
results are illustrated by applying the proposed algorithm on both
simulated and real benchmark datasets. Finally, discussion on the
choice of concentration parameter $\alpha$ for the Dirichlet
process prior is provided in Section \ref{discussion}. Conclusions
and future work are given as well.

\section{Methodology}\label{model}
\subsection{The finite mixtures of multivariate Normal-inverse Gaussian distributions}\label{model_mnig}
MNIG distribution is  a mean-variance mixture of a $d-$dimensional
multivariate Gaussian distribution with the inverse Gaussian
distribution such that \citep{BarndroffNielsen1997}:
\begin{eqnarray*}
	\Xv|U=u &\sim& \normal(\tilde\muv+u\Delta\tilde\betav,u\Delta),
	\\ U &\sim& \IG(\tilde\delta,\tilde\gamma)
\end{eqnarray*}
with the constraint that $|\Delta|=1$ for identifiability.  In
the above $\tilde\muv$ is $d \times 1$ vector of location
parameters, $\Delta$ is a $d \times d$ covariance matrix,
$\tilde\betav$ is a $d \times 1$ vector of skewness parameters,
for $\tilde\betav=0$ we get a symmetric variance mixture model and
$\tilde\delta,\tilde\gamma$ are the parameters of the Inverse
Gaussian mixing distribution.

\cite{protassov2004based} proposed an alternative
re-parameterization of MNIG distribution such that:
\begin{equation*}
\muv=\tilde\muv, \quad \gamma=\tilde{\gamma}\tilde{\delta}, \quad \Sigmav = \tilde{\delta}^2\Delta, \quad \betav=\tilde{\betav}\Sigmav.
\end{equation*}
Hence, the mean-variance mixture is of the following form:
\begin{eqnarray*}
	\Xv|U=u &\sim& \normal(\muv+u\betav,u\Sigmav), \\
	U &\sim& \IG(1,\gamma),
\end{eqnarray*}
where now $\Sigmav$ is not restricted and deriving maximum
likelihood estimates of the parameters is simplified. Under this
configuration, the density of MNIG distribution has the following
form:
\begin{equation}\label{MNIGden}
f_\Xv(\xv) =
\frac{1}{2^{\frac{d-1}{2}}}\left[\frac{\alpha}{\pi
	q(\xv)}\right]^{\frac{d+1}{2}}\exp\left({p(\xv)}\right)~K_{\frac{d+1}{2}}(\alpha
q(\xv))
\end{equation}
where,
\begin{equation*}
\alpha = \sqrt{\gamma^2 + \betav^\top\Sigmav^{-1}\betav},\quad p(\xv) = \gamma + (\xv - \muv)^\top\Sigmav^{-1}\betav,\quad q(\xv) = \sqrt{1 + (\xv-\muv)^\top \Sigmav^{-1}(\xv-\muv)}
\end{equation*}
and $K_\alpha(x)$ is the modified Bessel function of the third
kind of order $\alpha$ evaluated at $x$. The MNIG distribution not
only can capture the skewness of the data by the parameter
$\betav$, but also is able to accommodate heavier tails with a
smaller value of the parameter $\alpha$.  Finally the density
of the Inverse Gaussian distribution denoted as $\IG(1,\gamma)$ is
given as
\[
f(u)=
\frac{1}{\sqrt{2\pi}}\exp(\gamma)u^{-3/2}\exp\left\{-\half\left(\frac{1}{u}+\gamma^2u\right)\right\},
~~\gamma>0
\]
and has  a unit mean. 

By combining the
conditional $d-$dimensional multivariate normal density of $\Xv|U
= u$ with the marginal density of U, the joint probability density
can be written as:
\begin{equation*}
\begin{split}
f(\xv,u) = &f(\xv|u)f(u)\\
= &(2\pi)^{-1/2}|u\Sigmav|^{-1/2}\exp\left\{-\half (\xv - \muv - u\betav)^\top(u\Sigmav)^{-1}(\xv - \muv - u\betav) \right\}\\
&\times \frac{1}{\sqrt{2\pi}}\exp(\gamma)u^{-3/2}\exp\left\{-\half\left(\frac{1}{u}+\gamma^2u\right)\right\}\\
\propto & u^{-\frac{d+3}{2}}|\Sigmav|^{-1/2}\exp\left\{-\half\left(\frac{1 }{u}+\gamma^2u-2\gamma\right) -\half (\xv - \muv - u\betav)^\top(u\Sigmav)^{-1}(\xv - \muv - u\betav)\right\}\\
\end{split}
\end{equation*}\par
This parameterization was utilized by \cite{Karlis2009} for model-based clustering using mixtures of MNIG distributions.

Consider a finite mixture of MNIG distribution with $G$ components
and density
\[
f_\Xv(\xv | \thetav) = \sum\limits_{g=1}^G \pi_g
\frac{1}{2^{\frac{d-1}{2}}}\left[\frac{\alpha_g}{\pi
	q_g(\xv)}\right]^{\frac{d+1}{2}}\exp\left({p_g(\xv)}\right)~K_{\frac{d+1}{2}}(\alpha
q_g(\xv))
\] where
\begin{equation*}
\alpha_g = \sqrt{\gamma_g^2 +
	\betav_g^\top\Sigmav_g^{-1}\betav_g},\quad p_g(\xv) = \gamma_g +
(\xv - \muv_g)^\top\Sigmav_g^{-1}\betav_g,\quad q_g(\xv) = \sqrt{1
	+ (\xv-\muv_g)^\top \Sigmav_g^{-1}(\xv-\muv_g)}
\end{equation*}
and $p_g$ are the mixing proportions such as $\pi_g>0$ and
$\sum_{g=1}^G \pi_g=1$.

Consider $N$ independent observations $\xv_1,\dots,\xv_N$ coming
from a mixture of $G-$component mixture  of MNIG distributions.
Augmenting the observed data with the unobserved
$u_{ig},z_{ig}$, $i=1,\ldots,N$ and $g=1,\ldots,G$ we can derive
the complete-data likelihood for a mixture of MNIG distributions
which is written as :
\begin{equation}\label{complik}
\begin{split}
L(\theta) \propto &\prod_{g=1}^{G}\prod_{i=1}^{N}\left[\pi_g u_{ig}^{-\frac{d+3}{2}}|\Sigmav_g|^{-\half} \exp\left\{-\half\left(u_{ig}^{-1}+\gamma^2u_{ig}-2\gamma_g\right)\right\}\right.\\
&\left.\times \exp\left\{-\half\left(\xv_{i} - \muv_{g} - u_{ig}\betav_{g})^\top(u_{ig}\Sigmav_g)^{-1}(\xv_i - \muv_{g} - u_{ig}\betav_{g}\right)\right\}\right]^{z_{ig}}\\
\end{split}
\end{equation}
where for each observation $\xv_i$, $\zv_i =
(z_{i1},\dots,z_{iG})$ is the component indicator variable of the
form being $z_{ig}=1$ if $\xv_i$ belongs to group $g$ and 0 if
not. Also the $u_{ig}$ are the unobserved mixing variables that
led to the MNIG model per observation and component.

Here, $\thetav_g = (\pi_g, \muv_g, \betav_g,\gamma_g, \Sigmav_g)$ denote the parameters related to the $g^{th}$ component and $g=1,\dots,G$. It can be shown that the complete data likelihood of mixtures of MNIG distributions has the form of an exponential family such that
\begin{equation*}
L(\thetav) \propto \prod_{g =
	1}^{G}\left\{[r(\thetav_g)]^{t_{0g}}\cdot \prod_{i =
	1}^{N}[h(\xv_i,u_{ig})]^{z_{ig}}\times\exp Tr\left\{\sum_{j =
	1}^{5}\phi_{j}(\thetav_g)\tv_{jg}(\xv,\uv_{g})\right\}\right\},
\end{equation*}
where, $h(\cdot,\cdot)$ is a normalizing constant and the
component-specified functions for the parameters,
$\phi_{j}(\thetav_g)$, and the sufficient statistics 
$\tv_{jg} (\xv,\uv_{g})$, for
$j = 1,\dots,5$, are given as follows:
\begin{align*}
\phi_{1} (\thetav_g)&= \Sigmav_g^{-1}\betav_g, &\tv_{1g} (\xv,\uv_{g})&= \sumn z_{ig}\xv_i^{\top};\\
\phi_{2} (\thetav_g) &= \Sigmav_g^{-1}\muv_g, &\tv_{2g} (\xv,\uv_{g})&= \sumn z_{ig}u_{ig}^{-1}\xv_{i}^\top;\\
\phi_{3} (\thetav_g)&= -\half (\Sigmav_g^{-1}\betav_g\betav_g^\top + \frac{\gamma_g^2}{d}\Id_{d}), &t_{3g}(\xv,\uv_{g}) &= \sumn z_{ig}u_{ig};\\
\phi_{4} (\thetav_g)&= -\half (\Sigmav_g^{-1}\muv_g\muv_g^\top + \frac{1}{d}\Id_{d}), &t_{4g} (\xv,\uv_{g}) &= \sumn z_{ig}u_{ig}^{-1};\\
\phi_{5} (\thetav_g)&= -\half \Sigmav_g^{-1}, &\tv_{5g} (\xv,\uv_{g}) &= \sumn z_{ig} u_{ig}^{-1}\xv_i\xv_i^\top
\end{align*}
and $t_{0g} = \sumn{z_{ig}}$; where $\Id_{d}$ is the identity matrix of dimensions $d \times d$.
Therefore, conjugate priors could be assigned for the parameters,
and Gibbs sampling scheme can be utilized for parameter estimation
and clustering. If the conjugate prior distribution of $\thetav_g$
is of the form
\begin{equation*}
h(\thetav_g)\propto r(\thetav_g)^{a_{0,g}^{(0)}} \exp\left\{\sum_{j=1}^5\phi_{jg}(\thetav_g) {a_{j,g}^{(0)}}\right\},
\end{equation*}
with hyperparameters taking initial values $(a_{0,g}^{(0)},a_{1,g}^{(0)},\ldots,a_{5,g}^{(0)})$, then the posterior distribution is of the form
\begin{equation*}
h(\thetav_g\mid \xv)\propto r(\thetav_g)^{a_{0,g}} \exp\left\{\sum_{j=1}^5\phi_{jg}(\thetav_g) {a_{j,g}}\right\},
\end{equation*}
with
\begin{align*}
a_{0,g}&=a_{0,g}^{(0)}+t_{0g},\\
\av_{1,g}&=\av_{1,g}^{(0)}+\tv_{1g} (\xv,\uv_{g}),\\
\av_{2,g}&=\av_{2,g}^{(0)}+\tv_{2g} (\xv,\uv_{g}),\\
a_{3,g}&=a_{3,g}^{(0)}+t_{3g} (\xv,\uv_{g}),\\
a_{4,g}&=a_{4,g}^{(0)}+t_{4g} (\xv,\uv_{g}), ~\text{and}\\
\av_{5,g}&=\av_{5,g}^{(0)}+\tv_{5g} (\xv,\uv_{g}).\\
\end{align*}

\subsection{The Dirichlet process mixture (DPM) model}\label{model_dp}
One of the most commonly used applications of Dirichlet process
(DP)\citep{ferguson1973} is to  define a Dirichlet process prior
to a mixing measure \citep{muller2013}. Given $N$ independent
observations $\xv_i,\dots,\xv_N$, one can consider a model that
describes $\xv_{i}$'s as independent draws from a mixture of
distributions of the form $F(\theta)$, i.e. as a finite mixture
model, but with the mixing distributions as realizations of a
Dirichlet process with the mass or concentration parameter
$\alpha$ and a base distribution $P_0$. This model, known as the
Dirichlet process mixture (DPM) model, was proposed by
\cite{antoniak1974}, and are often expressed as the following
hierarchical model:
\begin{align*}
\xv_i \mid \theta_i &\sim F(\theta_i)\\
\theta_i &\sim P(\cdot)\\
P &\sim DP(\alpha,P_0).
\end{align*}
The above Dirichlet process models are also referred as the infinite mixture models for the reason that equivalent models can be derived by letting the number of component $G$ goes to infinity from finite mixture models with the following hierarchical structure \citep{neal2000,rasmussen2000}:
\begin{align*}
\xv_i \mid c_i,\thetav&\sim F(\thetav_{c_i})\\
c_i|\piv &\sim \text{Discrete}(\pi_1,\dots,\pi_G)\\
\thetav_g &\sim P_0\\
\piv &\sim Dir(\alpha/G,\dots,\alpha/G);
\end{align*}
where $c_i$ are the class indicators such that $c_i=g$ means that
observation $\xv_i$ comes from the $g^{th}$ cluster (i.e., $z_{ig}
= 1$), $\thetav = (\thetav_1,\dots,\thetav_G)$ is the collection
of $\theta_{c_i}$ with the later one represent the parameters that
describe the distribution of observations from class $c_i$, and
$\piv=(\pi_1,\dots,\pi_G)$ are the mixing proportions given a
symmetric Dirichlet prior. By integrating out $\piv$ and then
taking the limit as the number of  components $G$ approaches
infinity, the conditional prior of the class indicator $c_i$ is
\begin{equation}\label{puprior}
\begin{array}{lr}
p(c_i=g|\cv_{-i},\alpha) = \dfrac{n_{-i,g}}{N-1+\alpha},\\
p(c_i\neq c_j \text{ for all } j\neq i|\cv_{-i},\alpha)  = \dfrac{\alpha}{N-1+\alpha} .\\
\end{array}
\end{equation}
Here, $n_{-i,g}>0$ denotes the number of observations in cluster
$g$ except for the $i^{th}$ observation.
This is known as the Polya urn model. \cite{blackwell1973} proposed the urn scheme of constructing a realization of the Dirichlet  process.
\cite{neal2000}, employed the Polya urn formulation of the Dirichlet process and proposed a Gibbs sampling scheme for sampling from the
posterior distributions for a Dirichlet mixture model when conjugate priors are used. The Polya urn representation and the Gibbs sampling
framework has also been used by \cite{rasmussen2000} for the infinite mixture models framework to Gaussian mixture models
and \cite{sivaganesan2002} developed a clustering procedure based on the infinite Gaussian mixture model for clustering
microarray gene expression data where posterior distributions are estimated through Gibbs sampling.
An alternate construction of a Dirichlet Process can be done via
the stick-breaking construction, proposed by \cite{sethurama1994}.
In our work, we utilize the Polya urn scheme and develop a
Dirichlet process mixture model of MNIG.

\subsection{Dirichlet process mixture of multivariate normal inverse-Gaussian distributions}\label{model_dpmnig}
As stated by \cite{west1992}, the Dirichlet process mixture model
can be adopted by  any distributions that can have a
representation as an exponential family form. We have shown in
section \ref{model_mnig} that the density of MNIG distributions
has the exponential family form. Therefore, the Dirichlet process
mixture of MNIG distributions can be obtained by letting the
number of component $G \to \infty$ for a finite $G-$component
mixtures with the hierarchical structure as follows:
\begin{enumerate}
	\item Bottom Layer: Assume each of the observed data $\xv_1,\dots,\xv_N$ conditional on the  parameters
	$\thetav_g=(\muv_g,\betav_g,\gamma_g,\Sigmav_g),$ for $g = 1,\dots,G$, and the unobserved class label variables associated with this
	observation $\cv = (c_1,\dots,c_N)$ is a sample from an MNIG distribution:
	\begin{equation*}
	\xv_i|c_i=g,\thetav\sim MNIG(\thetav_g),\quad \thetav_g=(\muv_g,\betav_g,\gamma_g,\Sigmav_g),\quad   
	\end{equation*}
	\item Second Layer: Prior distributions for the parameters $\thetav_1,\dots,\thetav_G$ and the class label variables $\cv = (c_1,\dots,c_N)$
	are assigned as the following:
	\begin{enumerate}
		\item The prior distribution for  class label variables $\cv=(c_1,\ldots,c_N)$ are:    
		\begin{equation*}
		p(c_i\mid \pi_1,\ldots,\pi_G) =\prod_{g=1}^G \pi_g^{\mathbf{I}(c_i=g)}
		\end{equation*}
		where $\mathbf{I}(c_i=g)=1$ whenever $c_i=g$ and 0 otherwise.
		\item Conjugate priors are given to the parameters $\thetav_g = (\muv_g,\betav_g,\gamma_g,\Sigmav_g)$ based on
		common hyperparameters $\left\{a_{0}^{(0)}, \av_{1}^{(0)}, \av_{2}^{(0)}, a_{3}^{(0)}, a_{4}^{(0)}, \av_5^{(0)}\right\}$
		among all components:
		\begin{equation*}
		\begin{split}
		\gamma_g &\sim \normal\left(\dfrac{a_{0}^{(0)}}{a_{3}^{(0)}},\dfrac{1}{a_{3}^{(0)}}\right)\cdot\1v\left(\gamma_g > 0\right);\\
		\Sigmav_g^{-1}&\sim Wishart\left(a_0^{(0)},{\av_{5}^{(0)}}^{-1}\right);\\
		\left.\begin{pmatrix} \muv_g\\\betav_g\end{pmatrix}\right\vert\Sigmav_g^{-1}&\sim\normal\left[\begin{pmatrix} \muv_0^{(0)}\\ \betav_0^{(0)}\end{pmatrix},\begin{pmatrix}
		\tau_\mu^{(0)}\Sigmav_g^{-1} & \tau_{\mu\beta}^{(0)}\Sigmav_g^{-1}\\
		\tau_{\mu\beta}^{(0)}\Sigmav_g^{-1} & \tau_\beta^{(0)}\Sigmav_g^{-1}
		\end{pmatrix}\right].
		\end{split}
		\end{equation*}
		Here,
		\begin{align*}
		\muv_0^{(0)} &= \frac{a_{3}^{(0)}\av_{2}^{(0)} - a_{0}^{(0)}\av_{1}^{(0)}}{a_{3}^{(0)}a_{4}^{(0)}-{a_{0}^{(0)}}^2}, \quad
		\betav_0^{(0)} = \frac{a_{4}^{(0)}\av_{1}^{(0)} - a_{0}^{(0)}\av_{2}^{(0)}}{a_{3}^{(0)}a_{4}^{(0)}-{a_{0}^{(0)}}^2}, \\
		\tau_\mu^{(0)} &= a_{4}^{(0)}, \quad
		\tau_\beta^{(0)} = a_{3}^{(0)},\quad \text{and} \quad
		\tau_{\mu\beta}^{(0)} = a_{0}^{(0)}.
		\end{align*}
	\end{enumerate}
	\item Top Layer: Priors are assigned to the hyperparameters defined in the second layer such as:
	\begin{enumerate}
		\item Symmetric Dirichlet prior distributions are given for the mixing proportions $(\pi_1,\dots,\pi_G)$:
		\begin{equation*}
		(\pi_1,\dots,\pi_G\mid\alpha,G) \sim Dirichlet(\alpha/G,\dots,\alpha/G).
		\end{equation*}
		Here, $\alpha$ is set equals to $1$ in our framework.
		\item Prior distributions for the common hyperparameters $\left\{a_{0}^{(0)}, \av_{1}^{(0)}, \av_{2}^{(0)}, a_{3}^{(0)}, a_{4}^{(0)}, \av_5^{(0)}\right\}$ are given as:\
		\begin{equation}\label{hyperpriors}
		\begin{split}
		\av_{j}^{(0)} &\sim \normal(\cv_j,B_j)\text{ for }j = 1,2;\\
		a_{j}^{(0)} &\sim \text{Exp}(b_j)\text{ for }j = 0,3,4;\\
		\av_5^{(0)} &\sim Wishart(\nu_0,\Lambdav_0);
		\end{split}
		\end{equation}
		where $\text{Exp}(b_j)$ is an exponential distribution with a rate parameter $b_j$; and hence density function of a $a_{j}^{(0)}$ is given as the follows
		\begin{equation*}
		f(a_{j}^{(0)}) = b_j\exp\left\{-b_ja_{j}^{(0)}\right\}, j=0,3,4.
		\end{equation*}
		In addition, we set
		\begin{equation*}
		b_0 = 1/N;\quad b_3 = 1/\sumn{u_i};\quad b_4 = 1/\sumn{u_i^{-1}};
		\end{equation*}
		\begin{equation}\label{hyperhyper}
		\begin{array}{ll}
		\cv_1 = \sumn{\xv_{i}}, &\quad \quad \quad \quad B_1 = \Sigmav_{\xv};\\
		\cv_2 = \sumn{u_i^{-1}\xv_{i}}, &\quad \quad \quad \quad B_2 = \Sigmav_{u^{-1}\xv};\\
		\nu_0 = d+1; &\quad \quad \quad \quad  \Lambdav_0 = \Sigmav_{\xv}/\nu_0.
		\end{array}
		\end{equation}
		Here, $N$ is the total number of observation, $d$ is the dimension of the data, $\Sigmav_{\xv}$ and $\Sigmav_{u^{-1}\xv}$
		are the sample covariance matrices of the data $\xv_1,\dots,\xv_N$ and $u_1^{-1}\xv_1,\dots,u_N^{-1}\xv_N$ respectively.
	\end{enumerate}
\end{enumerate}

\subsection{Discussion on the choice of hyperparameters}
The concentration parameter $\alpha$ is known to have an effect on
the number  of components \citep{BDA3}. There are several
different ways of specifying $\alpha$ in the model. Fixing
$\alpha$ to a specific value, for example $1$, like we do in our
framework, is one of the most commonly used methods.
\cite{escobar1995} experimented with different $\alpha$ values on
Dirichlet process mixture of Gaussian distributions, observing a
result that smaller value of $\alpha$ will encourage the data to
group together. Both \cite{huelsenbeck2007} and \cite{onogi2011}
have observed that different $\alpha$ values could yield different
number of components when applying a Dirichlet process mixture of
Dirichlet distributions to specify population structure of genetic
data. \cite{west1992} pointed out that one can assign a Gamma
prior for $\alpha$ to gain flexibility. We experiment with
different values of $\alpha$ and also employ a Gamma prior for
$\alpha$ in our simulation studies, discussed in Section
\ref{discussion}. However, our results indicate that $\alpha = 1$
is doing a satisfactory work; yet using different values for
$\alpha$ or assigning a prior distribution on $\alpha$ did not
change the result.

Choice of the base distribution $P_0$ for the
parameters $\thetav$, in the second layer of the hierarchical
structure discussed above, is crucial to the clustering result as
well as it provides the prior information of the spread of data
inside each component \citep{BDA3,hejblum2019sequential}. As
suggested by \cite{BDA3}, flexibility of the selection of $P_0$
can be added by incorporating another layer of hyperparameters on
the parameters in $P_0$. This is what we have done in the top
layer part(b) of our model. Similar to both \cite{rasmussen2000}
and \cite{sivaganesan2002} in their infinite mixture model of
Gaussian distributions framework, we put data-driven priors on the
common hyperparameters $\left\{a_{0}^{(0)}, \av_{1}^{(0)},
\av_{2}^{(0)}, a_{3}^{(0)}, a_{4}^{(0)}, \av_5^{(0)}\right\}$, and
choose the third layer hyperparameters such that
$\left\{a_{0}^{(0)}, \av_{1}^{(0)}, \av_{2}^{(0)}, a_{3}^{(0)},
a_{4}^{(0)}, \av_5^{(0)}\right\}$ are centered around their expected
value when treating the data overall as one group. Details about
the calculation can be found in Appendix \ref{math_detail}.
Although observations ought not be utilized to define priors,
\cite{rasmussen2000} argued that this set of data-driven
parameters in the priors is equivalent to normalizing the
observations and performs similarly to unit priors in their
infinite mixture of Gaussian case.

\section{Posterior Distributions}\label{post}
\subsection{Posterior distribution for the class label variables}
Recall the Polya urn scheme, where it indicates that the
conditional prior in (\ref{puprior}) for the class label
variables, when taking the limit of $G\to \infty$, is proportional
to the number of observation associated with that component for
all existing components; or to $\alpha$ when it requires a new
component to be generated. Combine this conditional prior with the
complete-data likelihood in (\ref{complik}), one can calculate the
the conditional posterior of the class-label variable for the
observation $\xv_i$ is
\begin{subequations}\label{class_up}
	\begin{alignat}{2}
	&p(c_i =g|\cv_{-i},\xv_i,\thetav_g)=b\dfrac{n_{-i,g}}{N-1+\alpha}f_\Xv(\xv_i|\thetav_g), \label{class_up:old}\\
	&p(c_i\neq c_j \text{ for all } j\neq i|\cv_{-i},\xv_i,\thetav_g)
	=b\dfrac{\alpha}{N-1+\alpha}\nonumber \\
	&\times \int f_{\Xv}(
	\xv_i|\thetav_g) p(\thetav_g|a_{0}^{(0)}, \av_{1}^{(0)}, \av_{2}^{(0)}, a_{3}^{(0)}, a_{4}^{(0)},\av_5^{(0)})d\thetav_g, & \label{class_up:new}
	\end{alignat}
\end{subequations}
Here $f_{\Xv}( \xv_i|\thetav_g)$ is the density function of MNIG
distribution with parameter $\thetav_g =
(\muv_g,\betav_g,\gamma_g,\Sigmav_g)$ and $n_{-i,g}$ is the number
of observations that belongs to the $g^{th}$ component without
counting the $i^{th}$ observation; $b$ is a normalizing constant
to ensure all mixing proportions add up to $1$.

\subsection{Posterior distributions for the MNIG parameters}
As discussed in section \ref{model_dpmnig} for the second layer of
the Dirichlet  process mixture of MNIG distributions model, we use
conjugate priors and then the posterior distributions for each
parameter are given below:
\begin{equation*}
\begin{split}
\gamma_g|\cdot & \sim \normal\left(\dfrac{a_{0,g}}{a_{3,g}},\dfrac{1}{a_{3,g}}\right)\cdot\1v\left(\gamma_g > 0\right);\\
\Sigmav_g^{-1}|\muv_g,\betav_g,\cdot & \sim Wishart\left(a_{0,g},\Vv_{0,g}\right);\\
\left.\begin{pmatrix} \muv_g\\\betav_g\end{pmatrix}\right\vert\Sigmav_g^{-1},\cdot& \sim\normal\left[\begin{pmatrix} \muv_{0,g}\\ \betav_{0,g}\end{pmatrix},\begin{pmatrix}
\tau_{\mu,g}\Sigmav_g^{-1} & \tau_{\mu\beta,g}\Sigmav_g^{-1}\\
\tau_{\mu\beta,g}\Sigmav_g^{-1} & \tau_{\beta,g}\Sigmav_g^{-1}
\end{pmatrix}\right].
\end{split}
\end{equation*}
Using the prior distribution and likelihood,  we can show that the
group specific update of posterior of the hyperparameters
$a_{0,g},\av_{1,g},\av_{2,g},a_{3,g},a_{4,g}$ and $\av_{5,g}$ are
given as follows:
\begin{equation}\label{hyperup}
\begin{array}{ll}
a_{0,g}= a_{0}^{(0)} +  \sumn z_{ig},\quad & a_{3,g} = a_{3}^{(0)}+\sumn z_{ig}u_{ig},\\
\av_{1,g} = \av_{1}^{(0)} + \sumn z_{ig}\xv_i^{\top},\quad & a_{4,g} = a_{4}^{(0)}+\sumn z_{ig}u_{ig}^{-1},\\
\av_{2,g} = \av_{2}^{(0)} + \sumn z_{ig}u_{ig}^{-1}\xv_{i}^\top,\quad & \av_{5,g} = \av_{5}^{(0)} + \sumn{z_{ig}u_{ig}^{-1}\xv_{i}\xv_{i}^\top};\\
\end{array}
\end{equation}
and hence,
\begin{equation*}
\begin{split}
\Vv_{0,g}^{-1} =&\av_{5,g}+\muv_0^{(0)}\tau_{\mu}^{(0)}{\muv_0^{(0)}}^\top+\muv_0^{(0)}\tau_{\mu\beta}^{(0)}{\betav_0^{(0)}}^\top+\betav_0^{(0)}\tau_{\mu\beta}^{(0)}{\muv_0^{(0)}}^\top+\betav_0^{(0)}\tau_{\beta}^{(0)}{\betav_0^{(0)}}^\top\\
&-\left(\muv_{0,g}\tau_{\mu,g}\muv_{0,g}^\top+\muv_{0,g}\tau_{\mu\beta,g}\betav_{0,g}^\top+\betav_{0,g}\tau_{\mu\beta,g}\muv_{0,g}^\top+\betav_{0,g}\tau_{\beta,g}\betav_{0,g}^\top\right);
\end{split}
\end{equation*}
\begin{equation*}
\muv_{0,g} = \dfrac{a_{3,g}\av_{2,g} - a_{0,g}\av_{1,g}}{a_{3,g}a_{4,g}-{a_{0,g}}^2},\quad \betav_{0,g} = \dfrac{a_{4,g}\av_{1,g} - a_{0,g}\av_{2,g}}{a_{3,g}a_{4,g}-{a_{0,g}}^2},\quad \tau_{\mu,g} = a_{4,g},\quad \tau_{\beta,g} = a_{3,g}, \quad \tau_{\mu\beta,g} = a_{0,g}.
\end{equation*}
Detail on the derivations are presented in Appendix \ref{math_detail}.\par

For the latent variable $U$, which is essential for updating the
hyperparameters $a_{0,g},\dots,a_{4,g}$ in (\ref{hyperup}) for the
posteriors of the parameters $\theta_g =
(\muv_g,\betav_g,\gamma_g,\Sigmav_g)$, the posterior is conditional
on the observations $\xv_1,\dots,\xv_N$, and is a generalized
inverse-Gaussian (GIG) distribution:
\begin{equation*}
U_{ig}|\Xv = \xv_{i} \sim
\GIG\left(\dfrac{d+1}{2},q_{ig}^2(\xv_i),\alpha_g^2\right),
\end{equation*}
where,
\begin{equation*}
\alpha_g = \sqrt{\gamma_g^2 + \betav_g^\top\Sigmav_g^{-1}\betav_g},\quad \quad q_{ig}(\xv_i) = \sqrt{1 + (\xv_i-\muv_g)^\top \Sigmav_g^{-1}(\xv_i-\muv_g)}
\end{equation*}
Therefore, the conditional expectations of $U_{ig}$ and
$U_{ig}^{-1}$ given $\Xv=\xv_{i}$ are as follows:
\begin{equation}\label{up_u}
\begin{split}
\E\left(U_{ig}|\xv_i\right) &= \dfrac{q_{ig}(\xv_i)}{\alpha_g }\dfrac{K_{\lambda+1}\left(\alpha_g q_{ig}(\xv_{i})\right)}{K_{\lambda}\left(\alpha_g q(\xv_{i})\right)},\\ \E\left(U_{ig}^{-1}|\xv_i\right) &= \dfrac{\alpha_g}{q_{ig}(\xv_i)}\dfrac{ K_{\lambda-1}\left(\alpha_g q_{ig}(\xv_{i})\right)}{ K_{\lambda}\left(\alpha_g q(\xv_{i})\right)};
\end{split}
\end{equation}
where $\lambda = -\dfrac{d+1}{2}$.

\subsection{Posterior distributions for the hyperparameters}
Data-driven priors of the conjugate form are given to the common hyperparameters $\left\{a_{0}^{(0)}, \av_{1}^{(0)}, \av_{2}^{(0)},\right.\\ \left.a_{3}^{(0)}, a_{4}^{(0)}, \av_5^{(0)}\right\}$ and the resulting posterior distributions are as follows:
\begin{equation}\label{comhyp_up}
\begin{split}
a_0^{(0)}|(\thetav_1,\dots,\thetav_G) &\sim \text{Exp}\left(b_0 - \sum_{g=1}^{G}\left[\gamma_g-\muv_g^\top\Sigmav_g^{-1}\betav_g+\log\left(|\Sigmav_g|^{-\half}\right)+\log(\pi_g)\right]\right);\\
\av_{1}^{(0)}|(\thetav_1,\dots,\thetav_G) &\sim \normal\left(\cv_1+\sum_{g=1}^{G}\betav_g^\top\Sigmav_g^{-1} B_1, B_1\right);\\
\av_{2}^{(0)}|(\thetav_1,\dots,\thetav_G) &\sim \normal\left(\cv_2+\sum_{g=1}^{G}\muv_g^\top\Sigmav_g^{-1} B_2, B_2\right);\\
a_3^{(0)}|(\thetav_1,\dots,\thetav_G) &\sim \text{Exp}\left(b_3 +\half\sum_{g=1}^{G}\left(\betav_g^\top\Sigmav_g^{-1}\betav_g+\gamma_g^2\right)\right);\\
a_4^{(0)}|(\thetav_1,\dots,\thetav_G) &\sim \text{Exp}\left(b_4 +\half\sum_{g=1}^{G}\left(\muv_g^\top\Sigmav_g^{-1}\muv_g+1\right)\right);\\
\av_5^{(0)}|(\thetav_1,\dots,\thetav_G) &\sim
\text{Wishart}\left[\nu_0+G
a_0,\left(\Lambdav_0^{-1}+\sum_{g=1}^{G}\Sigmav_g^{-1}\right)^{-1}\right].
\end{split}
\end{equation}
See details on derivations in Appendix \ref{math_detail}.

\section{Posterior Estimation via Gibbs Sampling}\label{gibbs}
\subsection{A Gibbs sammpler}
Posterior estimation in a Bayesian framework can be  done by using
samples from the posterior distribution via Gibbs sampling.
\cite{BDA3} mentions that updating $c_i$ one
at a time when implementing the Gibbs sampling framework can often
leads to poor mixing. Therefore, we apply the ``marginal Gibbs
sampler'' as suggested by \cite{BDA3}, which update $\cv$ and
$\thetav_1,\dots,\thetav_G$ separately as two blocks, then iterate
between these two blocks. The detailed Gibbs sampler is presented
below:
\begin{enumerate}
	\item[Step 0] Initialization: The algorithm is initialized such that all observed data $\xv = (\xv_1,\dots,\xv_N)$ belong to one single group. In other words, we have $G=1$ and $c_1 = c_2 = \cdots= c_N$. Parameters of this one-component model are initialized as follows with $g=G=1$:
	\begin{enumerate}
		\item $\gamma_g = 1$.
		\item $\muv_g$ is set as the sample mean.
		\item $\betav_g$ is assigned a $d$-dimensional vector with all entries equal to $0.01$.
		\item $\Sigmav_g$ is initialized as the component sample variance matrix
		\item $\alpha_g$ and $q_{ig}(\xv_{i}), i=1,\dots,N$ are calculated based on the above parameters.
		\item $u_{ig}$ and $u_{ig}^{-1}$ are estimated using their conditional expected value from (\ref{up_u}), based on the parameters initialized as above.
		\item Data-driven hyperparameters $\left\{b_0,B_1,B_2,b_3,b_4,\cv_1,\cv_2,\nu_0,\Lambdav_0\right\}$ in the
		top layer of the model can be calculated based on (\ref{hyperhyper}).
		\item Common hyperparameters $\left\{a_{0}^{(0)}, \av_{1}^{(0)}, \av_{2}^{(0)}, a_{3}^{(0)}, a_{4}^{(0)}, \av_5^{(0)}\right\}$
		are initialized as sample averages of $10000$ samples from their prior distributions in (\ref{hyperpriors}).
	\end{enumerate}
	\item[Step 1] At $t^{th}$ iteration: Update $c_{i}$ from its conditional posterior distribution in (\ref{class_up}).
	Notice that it is infeasible that the integral in (\ref{class_up:new}) being evaluated analytically.
	Both \cite{neal2000} and \cite{rasmussen2000} suggest sampling from the prior of the parameters based
	on the common hyperparameters and implementing a Monte Carlo estimate to the probabilities.
	A new cluster is created when $c_i\neq c_j$ for all $i\neq j$ is selected and a cluster is removed if no observation is assigned to that
	cluster. After updating class label for all observations, the total number of component for the current iteration $G^{(t)}$ is also get updated.

	\item[Step 2]
	\begin{enumerate}
		\item For each of the current component: $\pi_g$ is set as the proportion of observation in the $g^{th}$ component.
		\item Update $\alpha_g$ and $q_{ig}(\xv_i)$ using the parameters carried from previous iteration $\theta_g^{(t-1)}$
		and the samples from common prior for the newly generated components, then compute $u_{ig}$ and $u_{ig}^{-1}$ based
		on their conditional expectation  from (\ref{up_u}).
		\item Based on the updated $u_{ig}$ and $c_{i}$, calculate the  group-specified hyperparameters\\ $\left\{a_{0,g},\av_{1,g},\av_{2,g},a_{3,g},a_{4,g},\av_{5,g}\right\}$, $g = 1,\dots, G^{(t)}$ using (\ref{hyperup}).
		\item Update the parameters to  $\thetav_g^{(t)}=(\muv_g,\betav_g,\delta_g,\gamma_g,\Sigmav_g)^{(t)}$ by each drawing one sample from their posterior distributions which depend on $\left\{a_{0,g},\av_{1,g},\av_{2,g},a_{3,g},a_{4,g},\av_{5,g}\right\}$, $g = 1,\dots, G^{(t)}$.
		\item Based on the current parameters $\thetav_g^{(t)}=(\muv_g,\betav_g,\delta_g,\gamma_g,\Sigmav_g)^{(t)}$, update the common hyperparameters to $\left\{a_{0,g},\av_{1,g},\av_{2,g},a_{3,g},a_{4,g},\av_{5,g}\right\}^{(t)}$ by drawing samples from their posterior distribution in (\ref{comhyp_up}).
	\end{enumerate}
\end{enumerate}
Step 1 and 2 are iterated until convergence.

\subsection{Convergence assessment and label switching issue}
Convergence is monitored using the potential scale reduction
factor \citep{gelman1992}, which is  based on the comparisons of
between and within variations among the different chains. To
generate these three likelihood chains, three independent
sequences initialized as:
\begin{enumerate}[noitemsep]
	\item all observations start in one group;
	\item each observation starts in its own group, hence $N$ different groups; and
	\item observations are clustered into $k-$ groups using randomly selected $k$ where $1\leq k \leq N$.
\end{enumerate}
Likelihood is calculated using the  updated parameters at the end
of each iteration. As early iterations reflect the starting
approximation and may not represent the target posterior, samples
from the early iterations known as ``burn-in'' period are
discarded \citep{BDA3}. The chains are considered as converged and
mixing well if the potential scale reduction factor calculated
based on the likelihood chains after ``burn-in'' is below 1.1.
Estimation of the parameters are done by averaging the samples
drawn from approximated posterior distributions after reaching a
stationary posterior distribution and discarding those from the
``burn-in'' period \citep{diebolt1994}. Here, we drew another 400
samples from the posterior distribution  for parameter estimation
after the potential scale reduction factor reaches below 1.1.\par

Label-switching issue, which is the invariance  of the likelihood
under permutation of the mixing components, can often occur in the
Bayesian approach to parameter estimation using mixture models
\citep{Stephens2000}. This leads multimodal posterior
distributions and inferences based on the posterior mean are not
appropriate \citep{Stephens2000}. \cite{Celeux2000} considered a
decision theoretic approach to overcome the label switching issue.
\cite{Stephens2000} proposed an algorithm that combined the
relabeling algorithm with the decision theoretic approach. An
alternate approach is to impose artificial constraints on a
certain set of parameters to force the labeling to be unique
\cite{richardson1997}. Constraints can be imposed on the mixing
proportions $\pi_1,\dots,\pi_G$ or any model parameters. In our
framework, we put constrains that the components are labeled such
that the first dimension of the location parameter $\mu_g$ follow
an ascending order which worked well in our case.

\subsection{Cluster allocation finalization and performance assessment}
To finalize the clustering result, we exploit the traditional
maximum {\it a-posteriori} probability (MAP) approach.  After
reaching a stationary approximation to the posterior distribution,
$400$ more samples are drawn, for each of the three sequences.
Within the $1200$ samples of class label variables, for each
observation, then we associate this observation with this most
frequently occurred component.  When the true class label is
known, the performance of the clustering algorithm can be assessed
using the adjusted Rand index \cite[ARI;][]{hubert1985}, which is
a measure of the pairwise agreement between the true and estimated
classifications after adjusting for agreement by chance. An ARI
value of `1' indicates a perfect agreement and a value of `0' is
expected for a classification with random guessing.

\section{Simulation Studies and Real Data Analysis}\label{results}
\subsection{Simulation Study 1}
\begin{figure}[h!]
	\begin{minipage}{.5\textwidth}
		\centering
		\includegraphics[width=.95\linewidth]{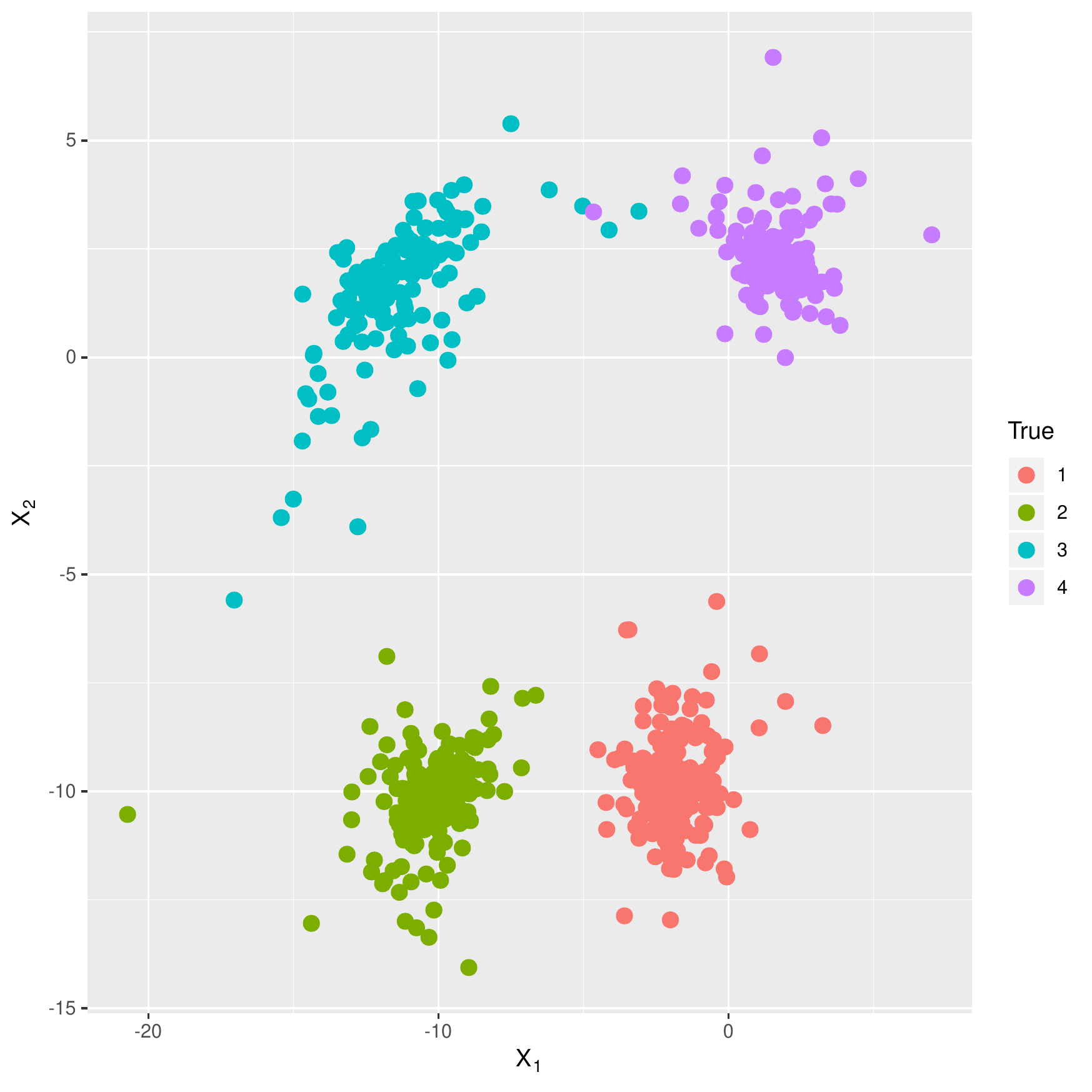}
	\end{minipage}%
	\begin{minipage}{.5\textwidth}
		\centering
		\includegraphics[width=.95\linewidth]{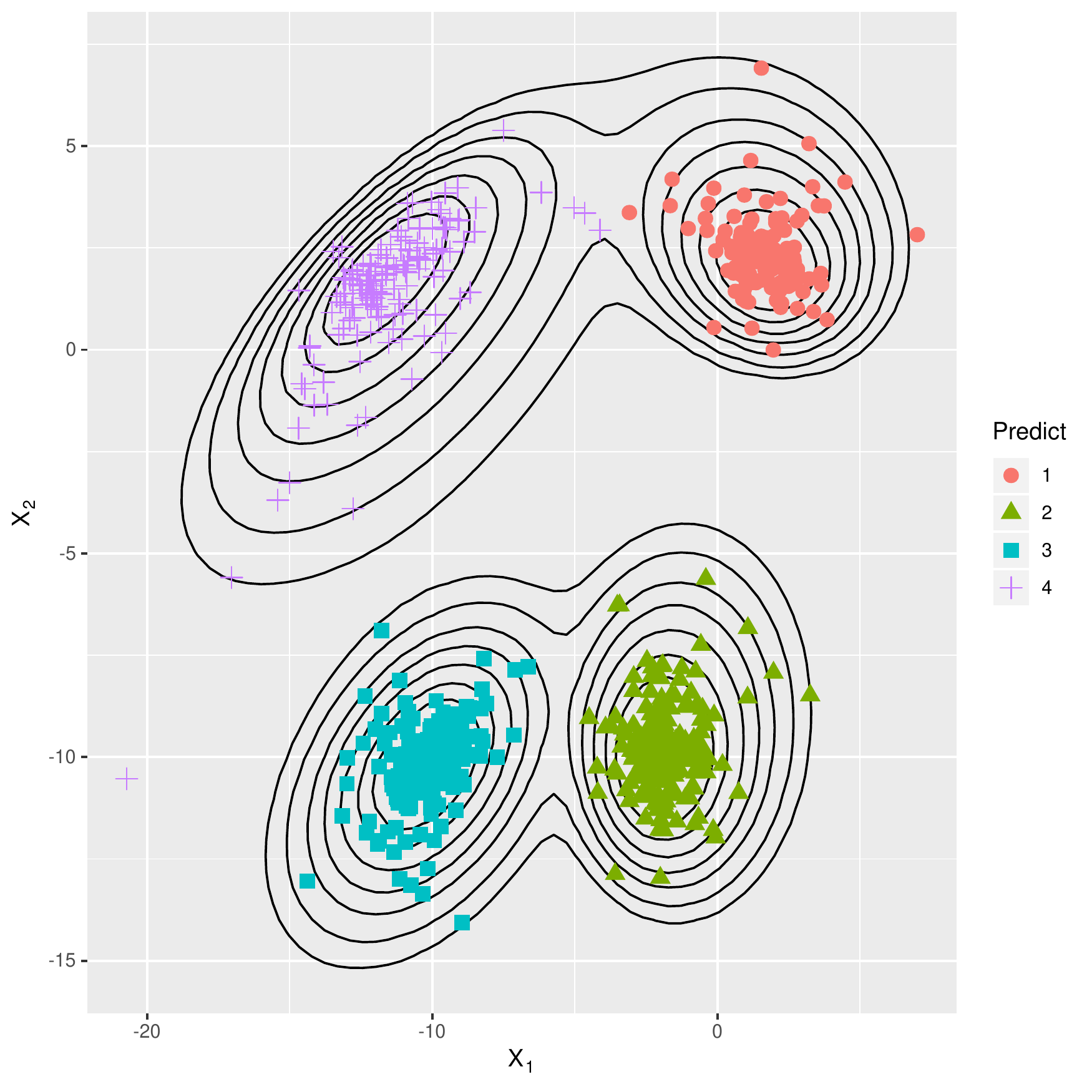}
	\end{minipage}
	\caption{Scatter plot highlighting the true labels (left) and a contour plot showing the predicted classifications (right). ARI = 0.989.}
	\label{fig:sim1}
\end{figure}
In this set of simulation study,  the proposed algorithm was
applied to 100 two-dimensional datasets, see Figure~\ref{fig:sim1}
(left panel) for one of the 100 datasets. Each dataset contained a
four components of skewed data with either heavier or lighter
tails, whose parameters are summarized in Table~\ref{tab:s1}. The
number of observations in each component were not balanced i.e.,
the four component comprised of $200,180,150,120$ observations
respectively. The proposed algorithm was applied to all 100
datasets. 100 out of 100 times it selected a four-component model
and the average ARI is $0.994$ (standard error of 0.001). The
estimated parameters are also summarized in
Table~\ref{tab:s1}.\par
\begin{table}[h!]
	\centering
	\caption{True and estimated values for the parameters in Simulation Study 1. Numbers in parentheses are the standard errors calculated from
		the replications}\label{tab:s1}
	\vspace*{0.1 in}
	\scalebox{0.75}{
		\renewcommand{\arraystretch}{1.75}
		\begin{tabular}{|c|c|c|c|c|}
			\hline  &\multicolumn{2}{|c|}{Component 1 ($n_1=200$)}&\multicolumn{2}{|c|}{Component 2 ($n_2=180$)}\\
			\hline  & True Parameters & Estimated Parameters & True Parameters & Estimated Parameters\\
			\hline  $\gamma$ & $1.2$ & $1.08 (0.33)$ & $0.8$ & $0.86 (0.29)$  \\
			\hline  $\muv$ & $[-2,\quad -10]$ &$[-2.00 (0.13),\quad -10.00 (0.12)]$ & $[-10,\quad -10]$ & $[-9.99 (0.16),\quad -10.03 (0.12)]$ \\
			\hline  $\betav$ & $[0.1,\quad 0.2]$ & $[0.09 (0.20),\quad 0.17 (0.20)]$ & $[-0.2,\quad -0.2]$ & $[-0.22 (0.21),\quad -0.18 (0.17)]$\\
			\hline   $\Sigmav$ &$\begin{bmatrix} 1.2&0\\0&1.2\end{bmatrix}$ & $\begin{bmatrix}
			1.11 (0.28) & 0.00 (0.09)\\0.00 (0.09)&1.09 (0.31)\end{bmatrix}$ &  $\begin{bmatrix} 1 & 0.4 \\ 0.4& 1\end{bmatrix}$ & $\begin{bmatrix}
			1.06(0.32) & 0.41(0.15) \\ 0.41(0.15) &1.03(0.28)\end{bmatrix}$ \\\hline
			Mean $\muv+\frac{\betav}{\gamma}$ & $[-1.92,\quad -9.83
			]$ & $[-1.92 (0.07),\quad -9.84(0.08)]$ & $[-10.25,\quad -10.25]$ & $[-10.23 (0.09),\quad -10.23(0.09)]$\\
			\hline
			Variance $\frac{\Sigmav}{\gamma}+\frac{\betav^\top\beta}{\gamma^3}$ & $\begin{bmatrix} 1.01&0.01\\0.01&1.02
			\end{bmatrix}$ & $\begin{bmatrix}
			1.11(0.20)&0.01(0.09)\\ 0.01(0.09)&1.11(0.18)\end{bmatrix}$ &
			$\begin{bmatrix}1.33&0.58\\0.58&1.33\end{bmatrix}$ & $\begin{bmatrix}
			1.43(0.26)& 0.59(0.16)\\0.59(0.16)&1.37(0.23)\end{bmatrix}$\\\hline
			\hline  &\multicolumn{2}{|c|}{Component 3 ($n_3=150$)}&\multicolumn{2}{|c|}{Component 4($n_4=120$)}\\
			\hline  & True Parameters & Estimated Parameters & True Parameters & Estimated Parameters\\
			\hline  $\gamma$ & $0.6$ & $0.65(0.21)$ & $1$ & $1.10(0.44)$  \\
			\hline  $\muv$ & $[-12,\quad 2]$ &$[-11.98(0.19),\quad 2.03(0.14)]$ & $[2,\quad 2]$ & $[2.00(0.18),\quad 2.01(0.15)]$ \\
			\hline  $\betav$ & $[0.2,\quad -0.25]$ & $[0.22(0.21),\quad -0.30(0.19)]$ & $[-0.2,\quad 0.2]$ & $[-0.26(0.37),\quad 0.23(0.28)]$\\
			\hline   $\Sigmav$ &$\begin{bmatrix} 2&1\\1&1\end{bmatrix}$ & $\begin{bmatrix}
			2.12(0.55) & 1.05(0.29)\\1.05(0.29)&1.04 (0.29)\end{bmatrix}$ &  $\begin{bmatrix} 1.2 & -0.2 \\ -0.2& 1\end{bmatrix}$ & $\begin{bmatrix}
			1.30(0.49) & -0.22(0.14) \\ -0.22(0.14) &1.07(0.41)\end{bmatrix}$ \\\hline
			Mean $\muv+\frac{\betav}{\gamma}$ & $[-11.67,\quad 1.58
			]$ & $[-11.64(0.17),\quad 1.57(0.12)]$ & $[1.80,\quad 2.20]$ & $[1.78 (0.12),\quad 2.21(0.10)]$\\
			\hline
			Variance $\frac{\Sigmav}{\gamma}+\frac{\betav^\top\beta}{\gamma^3}$ & $\begin{bmatrix} 3.52&1.44\\1.44&1.96\end{bmatrix}$ & $\begin{bmatrix}
			3.93(0.78)&1.55(0.36)\\1.55(0.36)&2.20(0.55)\end{bmatrix}$ &
			$\begin{bmatrix}1.24&-0.24\\-0.24&1.04\end{bmatrix}$ & $\begin{bmatrix}
			1.39(0.37)&-0.28(0.17)\\-0.28(0.17)&1.13(0.26)\end{bmatrix}$\\\hline
	\end{tabular}}
\end{table}
Figure~\ref{fig:sim1} (left) shows the true component  membership
of one of the hundred datasets and Figure~\ref{fig:sim1} (right)
gives the contour plot based on the estimated parameters for the
dataset described in the left panel. The 95\% credible intervals
for all parameter estimation for all 100 datasets are given in the
Appendix \ref{supp_fig}, where the lower and upper endpoints of
the credible intervals are the empirical 0.025-percentiles and
0.975-percentiles.\par

We compared our approach with  other commonly used mixture models:
Gaussian mixture models (GMM) implemented in the {\sf R} package
\texttt{mclust} \citep{mclust}, and mixtures of generalized
hyperbolic distributions \citep[MixGHD;][]{MGHD}, which also have
the flexibility of modeling skewed as well as symmetric
components, and implemented in the {\sf R} package \texttt{MixGHD}
\citep{MixGHD}. These mixture models were applied to all 100
datasets. Gaussian mixture models failed to capture the skewness
of the components and overestimated the number of components (i.e.
always selected a five or more component model) while the mixture
of generalized hyperbolic distributions selected a four-component
model only 77 out of 100 times and gave average ARI $=0.975$ with
a standard deviation (sd) of $0.037$.

\subsection{Simulation Study 2}
In this simulation, 100 four-dimensional datasets  were generated
with three underlying groups with two components comprising of 200
observations, and the third component comprising of 100
observations.  The parameters used to generate the data are
summarized in Table~\ref{tab:s2}. The proposed algorithm was
applied to these 100 datasets and it correctly selected the
correct three-component model for all 100 datasets with an average
ARI of 1.000 (sd of 0.001). \par
\begin{figure}[h!]
	\begin{minipage}{.5\textwidth}
		\centering
		\includegraphics[width=.95\linewidth,height=3 in]{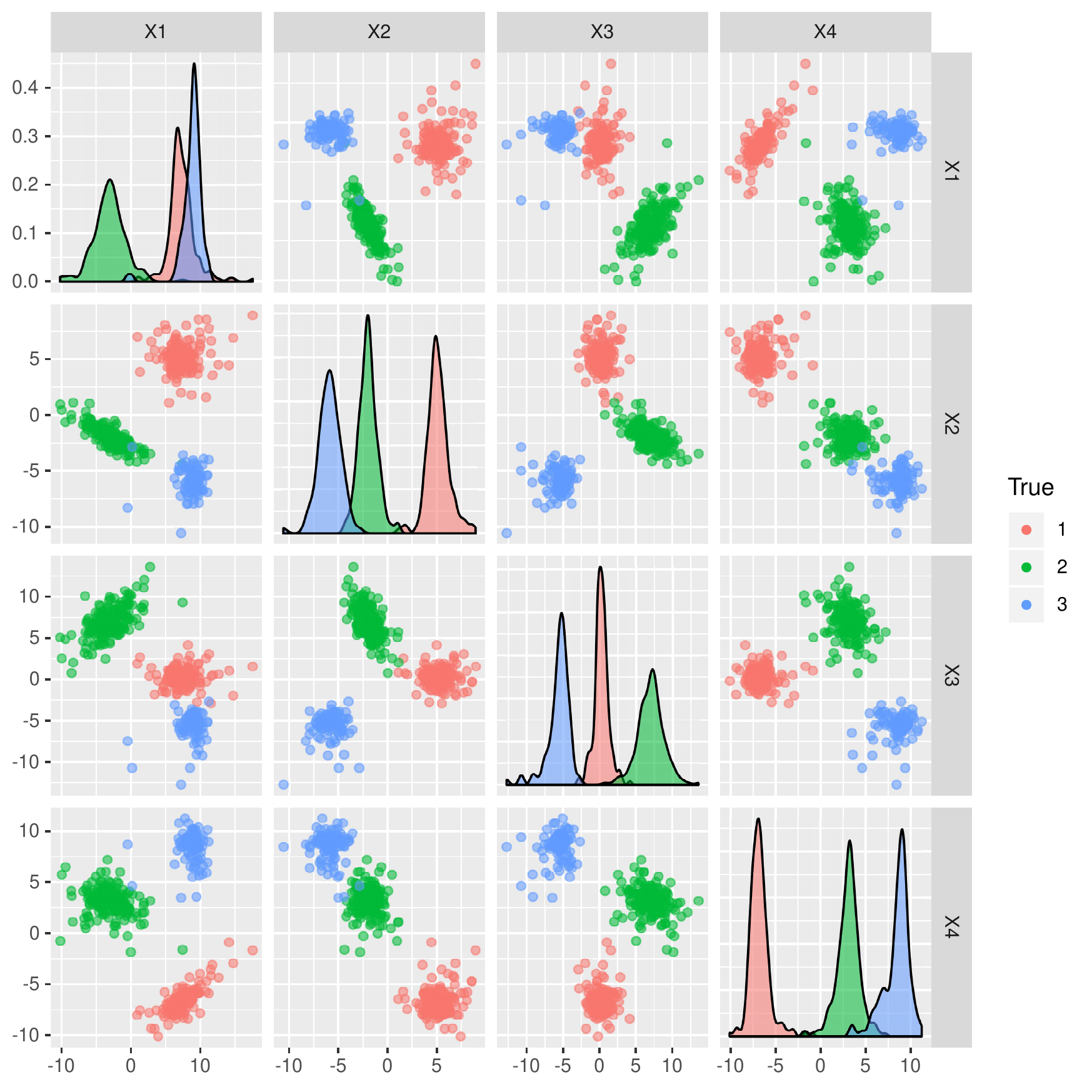}
	\end{minipage}%
	\begin{minipage}{.5\textwidth}
		\centering
		\includegraphics[width=.95\linewidth,height=3 in]{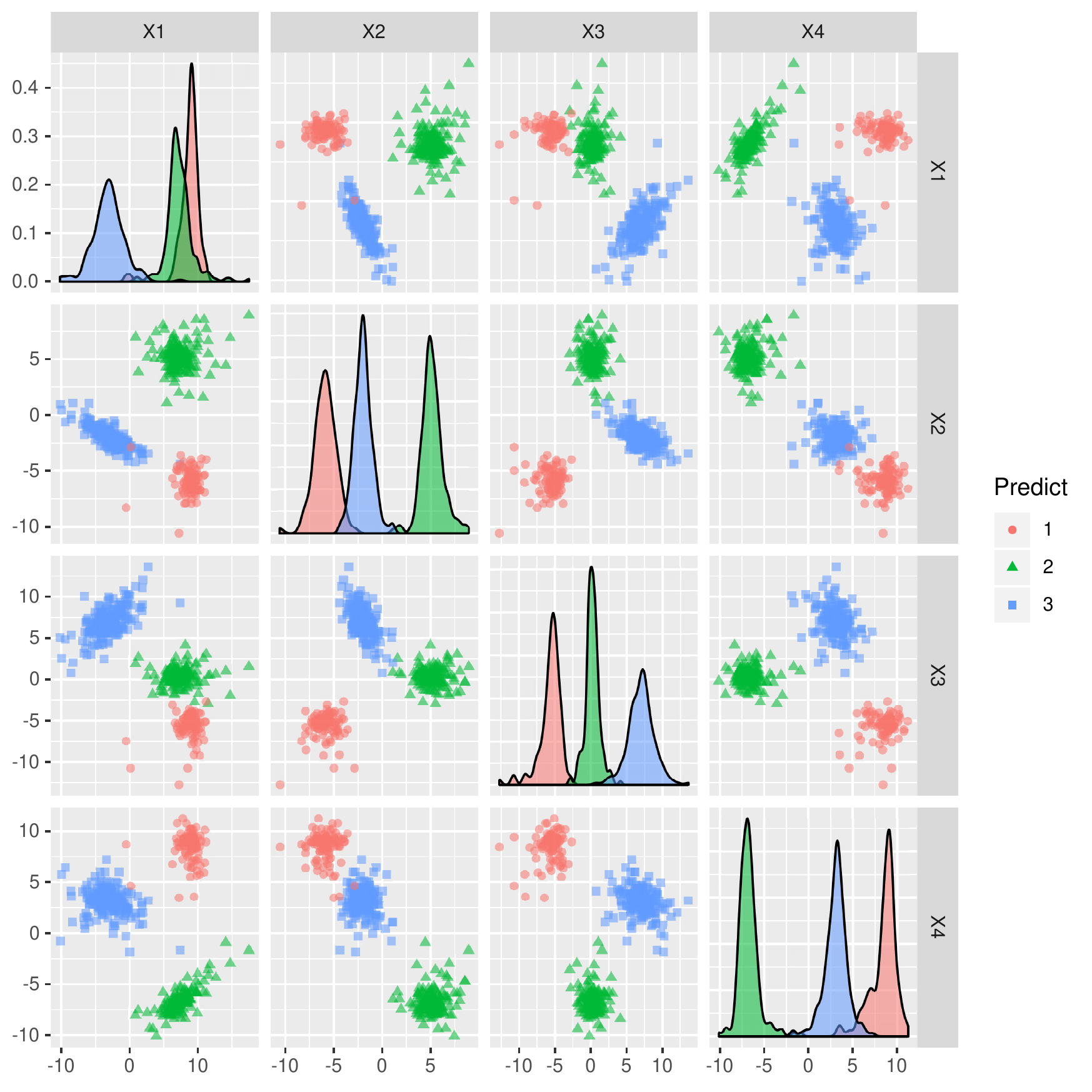}
	\end{minipage}
	\caption{Pairwise scatter plot highlighting the true labels (left) and predicted classifications (right) for one of the hundred datasets. Here, the ARI was 1.}
	\label{fig:sim2}
\end{figure}

\begin{table}[h!]
	\centering
	\caption{True and estimated values for the parameters in Simulation Study 2}\label{tab:s2}
	\vspace*{0.1 in}
	\scalebox{0.85}{
		\renewcommand{\arraystretch}{1.3}
		\begin{tabular}{|c|c|c|}
			\hline  &\multicolumn{2}{|c|}{\textbf{Component 1} ($n_1=100$)}\\
			\hline  & \textbf{True Parameters} & \textbf{Estimated Parameters (Standard Error)} \\
			\hline  $\gamma$ & $0.6$ & $0.79 (0.27)$  \\
			\hline  $\muv$ & $[9,-6,-5,9]$ & $[9.01(0.15), -6.00(0.16), -4.96(0.18), 9.04(0.21)]$ \\
			\hline  $\betav$ & $[0,0,-0.5,-0.5$ & $[0.00(0.21),0.01(0.22),-0.72(0.37),-0.72(0.40)]$ \\
			\hline   $\Sigmav$ & $\begin{bmatrix} 1&0&0&0\\0&1&0&0\\0&0&1&0\\0&0&0&1\end{bmatrix}$ & $\begin{bmatrix}  1.29 (0.42)& 0.02 (0.17)& 0.00 (0.15)&0.00 (0.17)\\0.02 (0.17)& 1.33 (0.46)& 0.02 (0.16)& 0.01 (0.18)\\0.00 (0.15)& 0.02 (0.16)&  1.28 (0.40)& -0.02 (0.17)\\ 0.00 (0.17)& 0.01 (0.18)& -0.02 (0.17)& 1.30 (0.45)\end{bmatrix}$  \\
			\hline
			\hline  &\multicolumn{2}{|c|}{\textbf{Component 2} ($n_2=200$)}\\
			\hline  & \textbf{True Parameters} & \textbf{Estimated Parameters (Standard Error)} \\
			\hline  $\gamma$ & $0.9$ & $0.98 (0.22)$  \\
			\hline  $\muv$ & $[7,5,0,-7]$ & $[6.99(0.15), 4.96(0.11), -0.02 (0.09), -7.00 (0.13)]$ \\
			\hline  $\betav$ &  $[0.2,0.2,0.2,0.2]$ & $[0.23(0.22),0.26(0.15),0.25(0.14),0.23(0.16)]$ \\
			\hline   $\Sigmav$ & $\begin{bmatrix} 2&0&0&1\\0&1&0&0\\0&0&1&0\\1&0&0&1\end{bmatrix}$ & $\begin{bmatrix} 2.19 (0.46)& -0.01 (0.11)& -0.02 (0.13)&1.10 (0.24)\\-0.01 (0.11)& 1.08 (0.21)& 0.00 (0.10)& -0.01 (0.09)\\-0.02 (0.13)& 0.00 (0.10)& 1.11 (0.23)& -0.02 (0.09)\\1.10 (0.24) & -0.01 (0.09) & -0.02 (0.09) & 1.11 (0.23)\end{bmatrix}$  \\
			\hline
			\hline  &\multicolumn{2}{|c|}{\textbf{Component 3} ($n_3=200$)}\\
			\hline  & \textbf{True Parameters} & \textbf{Estimated Parameters (Standard Error)} \\
			\hline  $\gamma$ & $1.2$ & $1.28 (0.28)$ \\
			\hline  $\muv$ & $[-3,-2,7,3]$ & $[-2.97(0.25), -2.00(0.11), 7.02(0.21), 3.00(0.16)]$\\
			\hline  $\betav$  & $[0,0,0,0]$ & $[-0.01(0.41),0.00(0.18),0.01(0.34),-0.02(0.25)]$\\
			\hline   $\Sigmav$ & $\begin{bmatrix} 6&-2&3&-1\\-2&1&-1&0\\3&-1&4&-1\\-1&0&-1&2\end{bmatrix}$ & $\begin{bmatrix}6.48(1.37)&-2.16(0.46)&3.23(0.76)&-1.03(0.37)\\ -2.16(0.46)& 1.10(0.28)& -1.08(0.27)& -0.02(0.13)\\3.23(0.76)& -1.08(0.27)& 4.32(0.97) &-1.06(0.39)\\  -1.03(0.37)& -0.02(0.13)& -1.06(0.39)& 2.17(0.51) \end{bmatrix}$ \\
			\hline
			
	\end{tabular}}
\end{table}
The average of the estimated parameters for all 100 datasets provided in Table~\ref{tab:s2} shows good parameter recovery. Figure~\ref{fig:sim2} (right) gives the pairwise scatter plot based on the estimated parameters for this dataset where the true group labels are described in the left panel.\par
Gaussian mixture models and mixtures of generalized hyperbolic distributions were applied to these datasets. The mixture of generalized hyperbolic distributions correctly selected a three-component model for 87 out of the 100 datasets with an average ARI  of $0.976$ (sd 0.066). The Gaussian mixture models only chose the correct number of components for 1 out of the 100 datasets.\par

\subsection{Real Data Analysis }
The proposed algorithm is also applied to some benchmark clustering datasets

\begin{figure}[h!]
	\begin{minipage}{.5\textwidth}
		\centering
		\includegraphics[width=.95\linewidth,height=3 in]{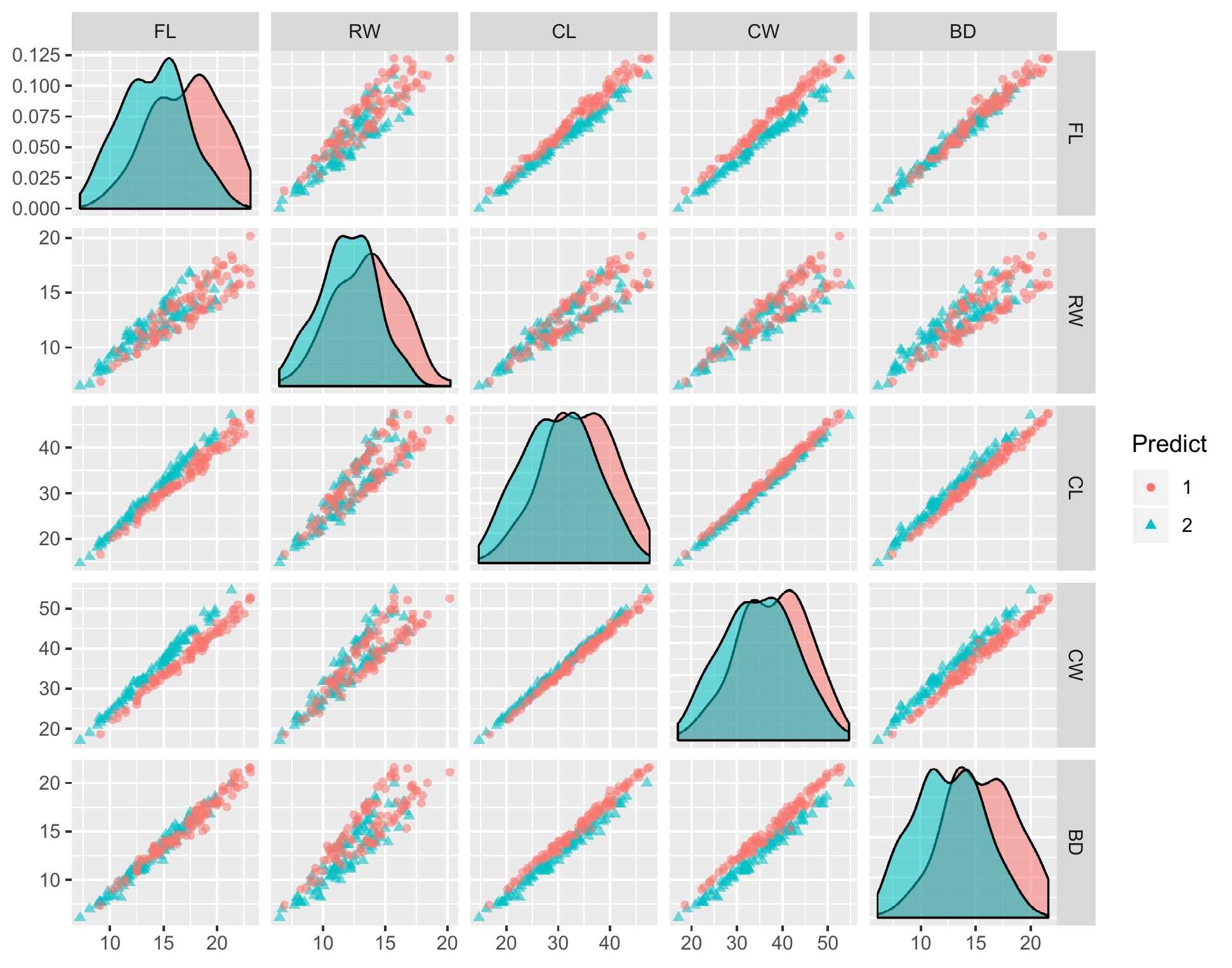}
	\end{minipage}%
	\begin{minipage}{.5\textwidth}
		\centering
		\includegraphics[width=.95\linewidth,height=3 in]{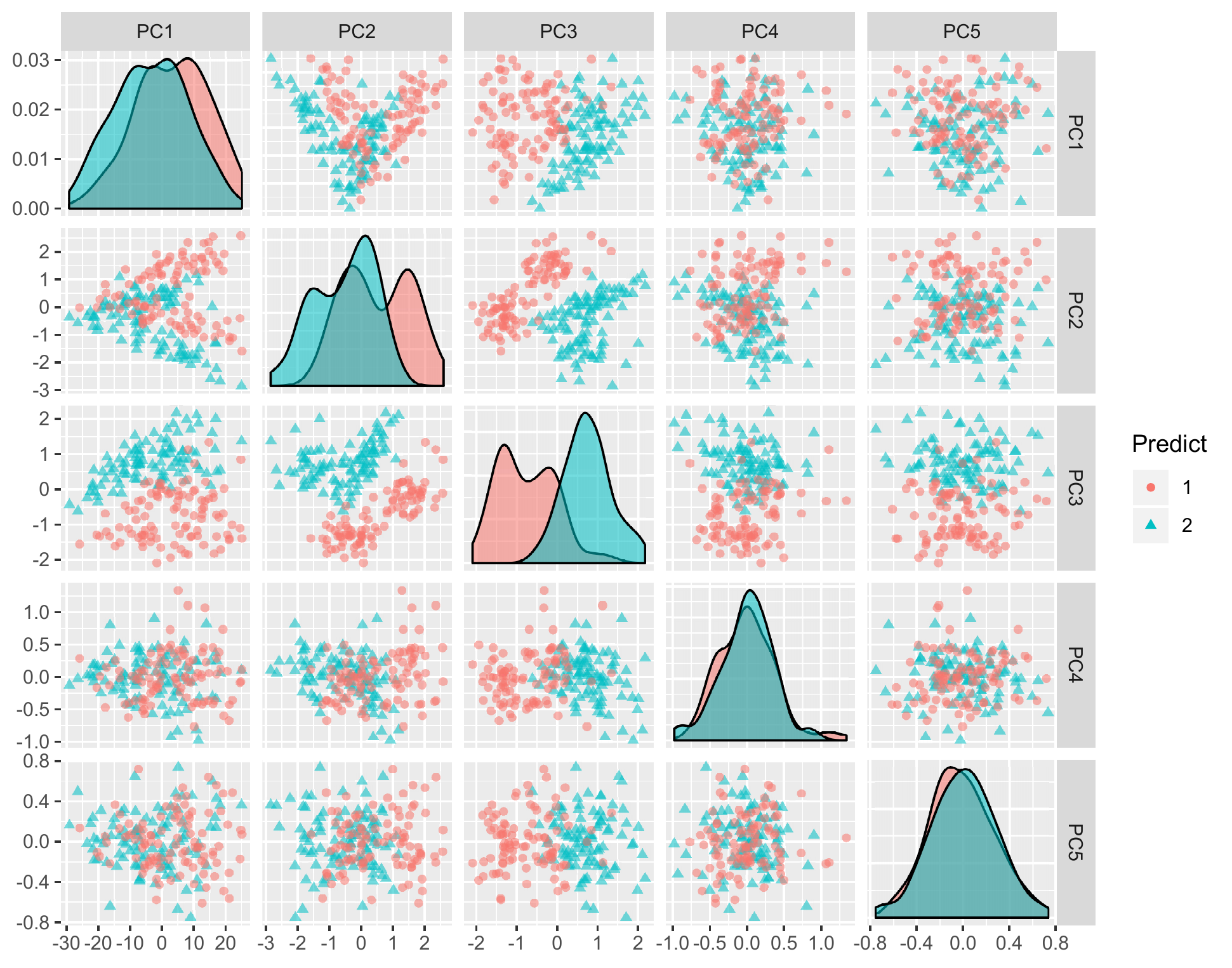}
	\end{minipage}
	\caption{Pairwise scatter plot highlighting the predicted classification of the color forms of the crabs, in terms of the original measurements (left), and in terms of the principal components (right).}
	\label{fig:crabs}
\end{figure}

\noindent \textbf{The Crab Dataset}:\\
This is dataset of morphological measurements on Leptograpsus crabs, available in the \texttt{R} package \texttt{MASS} \citep{MASS}. There are 200 observations and 5 covariates in this dataset, describing 5 morphological measurements on 50 crabs each of two colour forms and both sexes. The five measurements are frontal lobe size (FL), rear width (RW), carapace length (CL), carapace width (CW), and body depth (BD) respectively. All measurements are taken in the unit of millimeters. The proposed algorithm is applied to this dataset and it indicates a two-component model. Comparison of the estimated group membership with the two color forms of the crabs, ``B'' or ``O'' for blue or orange shows complete agreement (ARI$=1$). The pairwise scatter plots are given in Figure~\ref{fig:crabs}, where the left panel shows the original measurement variables and the right panel gives the principal components (only for visualization purposes), both colored with estimated classification of the color forms.\par
The Gaussian mixture models (GMM) and the mixtures of generalized
hyperbolic distributions (MixGHD) were also applied to this
dataset and the performance is summarized in
Table~\ref{tab:compare_crabs}. (IMMNIG denotes the infinite
mixture of MNIG developed here)
Both the mixtures of generalized hyperbolic distributions and the Gaussian mixture model select a three component model; the classification obtained by Gaussian mixture model are also more in agreement with the classification based on the gender of the crabs (ARI of 0.72) (see Table \ref{tab:compare_crabs}), whereas  the estimated group membership by the mixture of generalized hyperbolic distributions are more in agreement with classifying the crabs by their sexes (ARI of 0.51) (see Table \ref{tab:compare_crabs}).\\
\begin{table}[h!]
	\centering
	\vspace*{0.1 in}
	\caption{Summary of the performances of the new model (IMMNIG), the Gaussian mixture model (GMM), and mixtures of
		generalized hyperbolic distributions (MixGHD) on the crabs datasets.}
	\begin{tabular*}{\textwidth}{c@{\extracolsep{\fill}} ccc}
		\hline
		&   \textbf{Model Chosen}   &   \textbf{ARI (color)}    & \textbf{ARI (gender)}\\\hline
		IMMNIG & $G=2$ & 1.00 & 0.00 \\
		GMM &   $G=3$   &   0.02 & 0.70 \\
		MixGHD  &   $G=3$ & 0.25    &   0.51    \\\hline
	\end{tabular*}
	\label{tab:compare_crabs}
\end{table}

\begin{figure}[h!]
	\begin{minipage}{.5\textwidth}
		\centering
		\includegraphics[width=.95\linewidth,height=3 in]{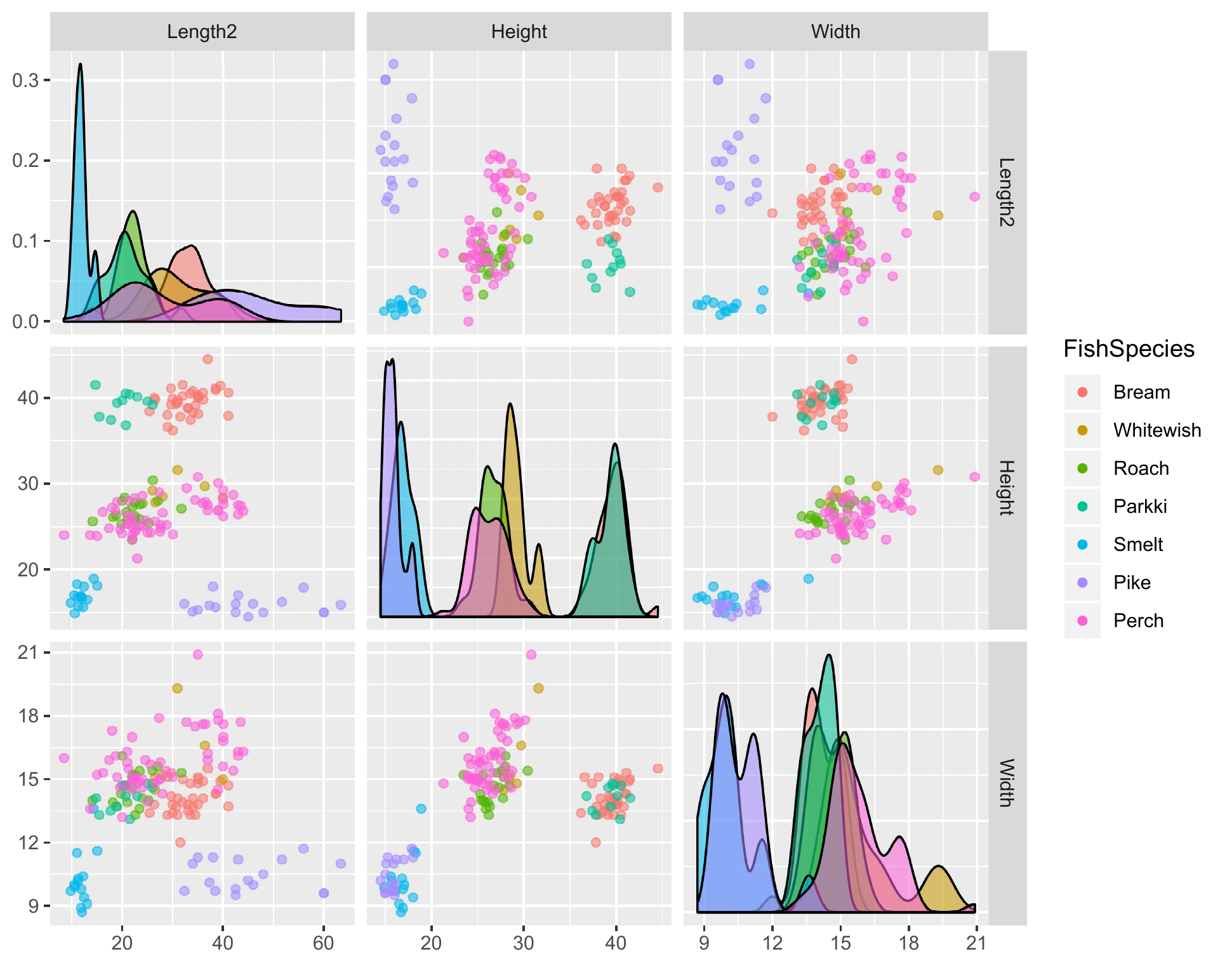}
	\end{minipage}%
	\begin{minipage}{.5\textwidth}
		\centering
		\includegraphics[width=.95\linewidth,height=3 in]{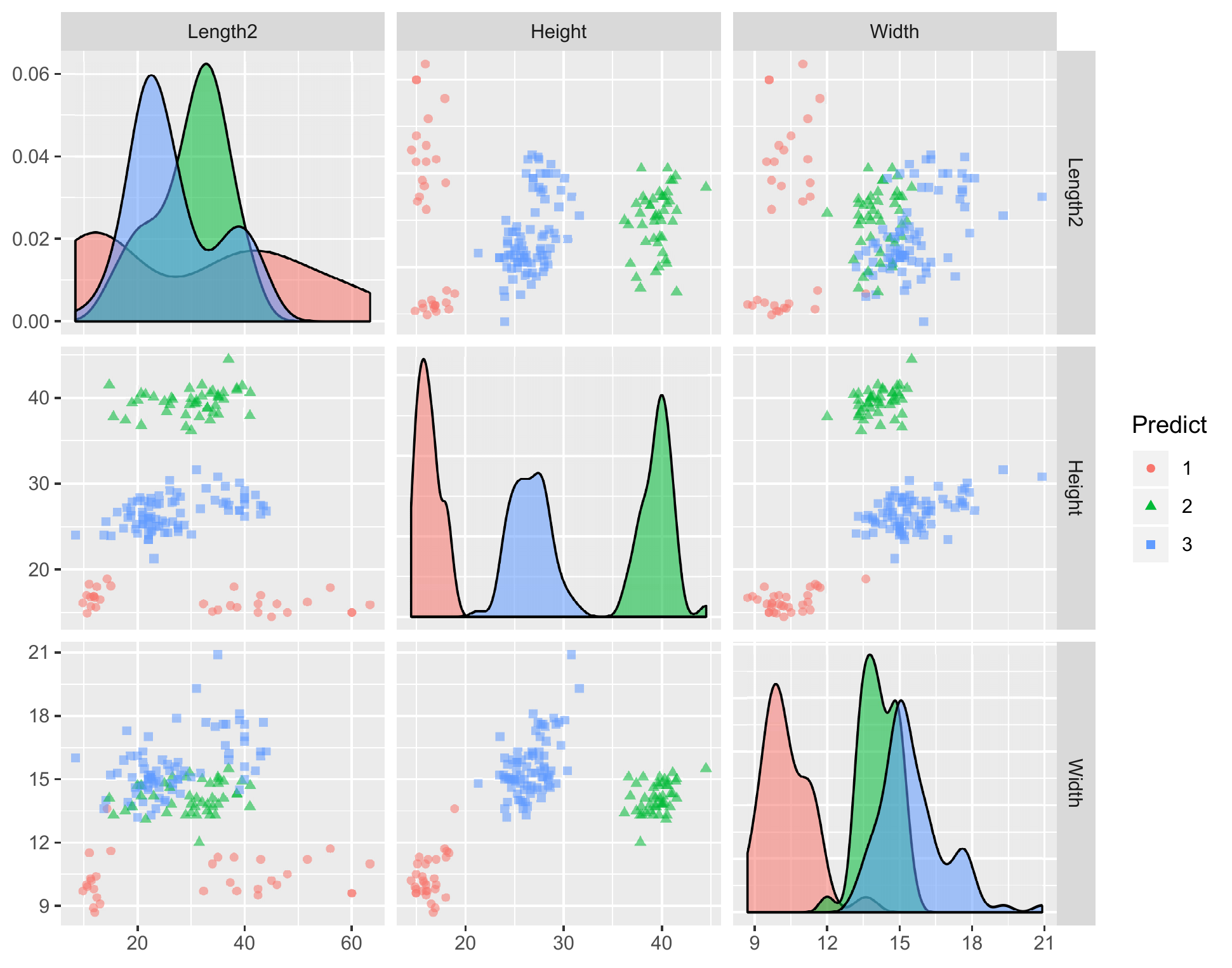}
	\end{minipage}
	\caption{Pairwise scatter plot highlighting the true labels (left) and a pairwise scatter plot showing the predicted classifications (right) for the fish catch data.}
	\label{fig:fishcatch}
\end{figure}

\noindent \textbf{The Fish Catch Dataset}\\
The fish catch data, available from the \texttt{R} package
\texttt{rrcov}, consists of the weight and different body lengths
measurements of seven different fish species. There are 159
observations in this data set. Similar to \cite{Subedi2014}, after
dropping the highly correlated variables, the variables
\texttt{Length2}, \texttt{Height} and  \texttt{Width} were used
for the analysis where \texttt{Length2} is the length from the
nose to the notch of the tail, \texttt{Height} is the maximal
height as a percentage of the length from the nose to the end of
the tail, and \texttt{Width} is the maximal width as a percentage
of the length from the nose to the end of the tail. The proposed
algorithm was applied (after scaling the data) and it selected a
three-component model. Figure~\ref{fig:fishcatch} shows the
pairwise scatter plots for this dataset, with the left panel
showing the true species of the fish and the right panel showing
the estimated component. Although the true number of species of
fish is seven, from the pairwise scatter plot, it is hard to
distinguish the species \texttt{White, Roach} and \texttt{Perch};
also, separation between the species \texttt{Bream} and
\texttt{Parkki}, as well as between \texttt{Smelt} and
\texttt{Pike} are not very clear. Table~\ref{tab:fishco}
summarizes the cross tabulation of the true species and estimated
group membership. The GMM and mixGHD are also applied to the
\texttt{Fish Catch} data and both resulted in a five component
model with classification where the additional fifth component
contained fish from  both \texttt{Whitewish} and \texttt{Perch}
(see Table \ref{tab:fishco} for detail). However, do note that
mixGHD was able to effectively separate \texttt{Smelt} and
\texttt{Pike}.

\vspace*{0.2 in}
\begin{table}
	\begin{center}
		\centering
		\renewcommand{\arraystretch}{1.3}
		\caption{Cross tabulation of true fish species and the estimated classification using the proposed model,
			Gaussian mixture models, and the mixtures of generalized hyperbolic distributions. }
		\begin{tabular*}{\textwidth}{c@{\extracolsep{\fill}} ccccc|ccccc|ccccc}
			\hline
			&\multicolumn{5}{c}{\textbf{IMMNIG}}&\multicolumn{5}{c}{\textbf{GMM}}&\multicolumn{5}{c}{\textbf{MixGHD}}\\
			&\multicolumn{5}{c}{ARI: 0.59}&\multicolumn{5}{c}{ARI: 0.52}&\multicolumn{5}{c}{ARI: 0.54}\\
			&\multicolumn{5}{c}{Estimated Groups}&\multicolumn{5}{c}{Estimated Groups}&\multicolumn{5}{c}{Estimated Groups}\\\hline
			&       &   1   &       &   2   &   3   &   1   &   2   &   3   &   4   &   5   &   1   &   2   &   3   &   4   &   5   \\\hline
			Bream   &       &   34  &       &       &       &   34  &       &       &       &       &   34  &       &       &       &       \\
			Parkki  &       &   11  &       &       &       &   11  &       &       &       &       &   11  &       &       &       &       \\\hline
			Whitewish   &       &       &       &   6   &       &       &   3   &   3   &       &       &       &   3   &   3   &       &       \\
			Roach   &       &       &       &   20  &       &       &   20  &       &       &       &       &   20  &       &       &       \\
			Perch   &       &       &       &   56  &       &       &   36  &   20  &       &       &       &   36  &   20  &       &       \\\hline
			Smelt   &       &       &       &       &   14  &       &       &       &   11  &   3   &       &       &       &   14  &       \\\hline
			Pike    &       &       &       &       &   17  &       &       &       &       &   17  &       &       &       &       &   17  \\\hline
		\end{tabular*}
		\label{tab:fishco}
	\end{center}
\end{table}

\begin{figure}
	\centering
	\includegraphics[scale=0.4]{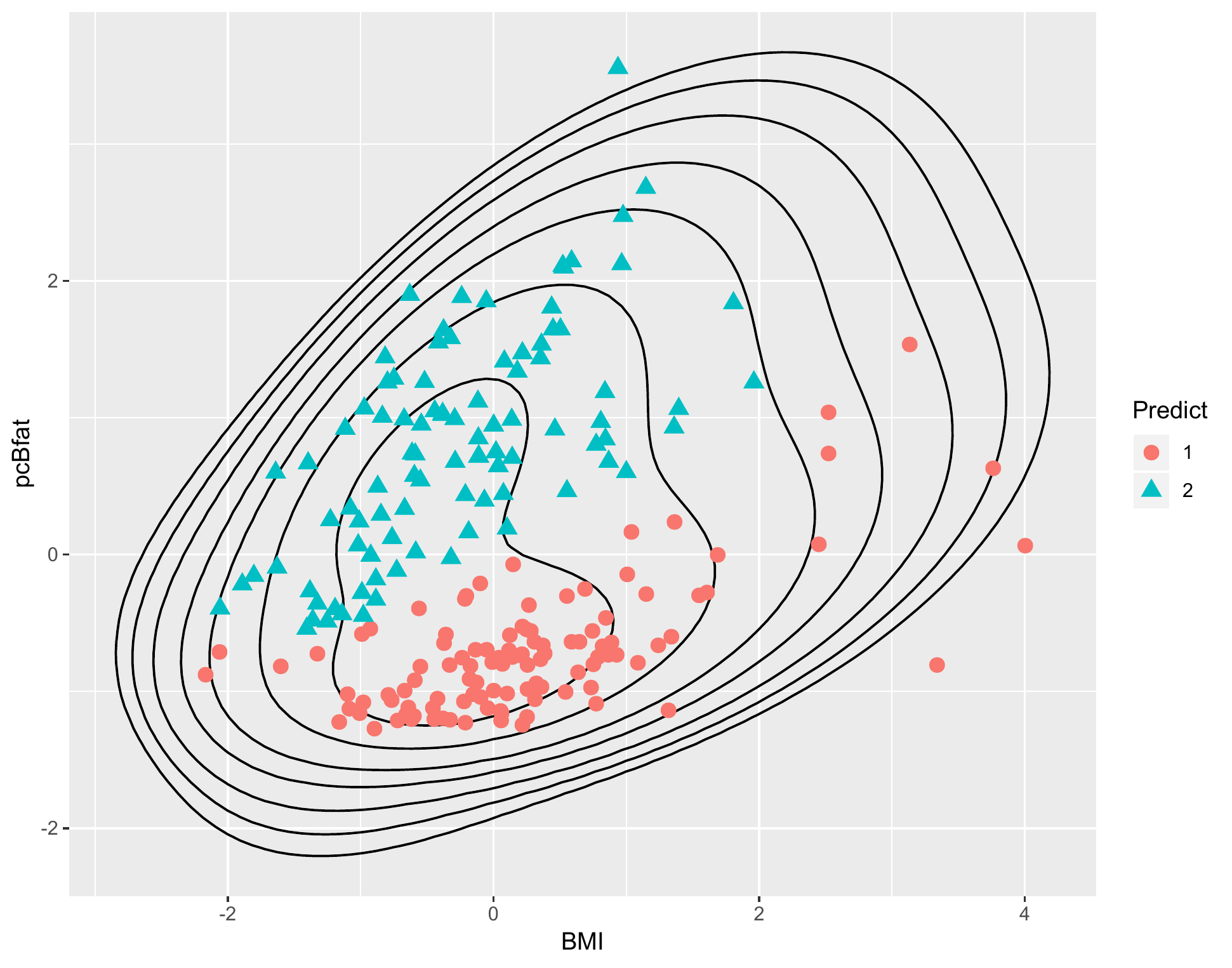}
	\caption{Contour plot using the estimated parameters for the AIS data}
	\label{fig:contor_ais}
\end{figure}

\noindent \textbf{The Australian Athletes (AIS) Dataset}:\\
The AIS dataset available in the \texttt{R} package \texttt{DAAG} \citep{DAAG} contains 202
observations and 13 variables comprising of measurements on various characteristics of the blood, body size, and sex of the athlete.
The proposed algorithm was applied on a subset of dataset with the variables body mass index (\texttt{BMI}) and body fat (\texttt{Bfat})
as it has been previously used \citep{vrbik2012,Lin2010}. The algorithm is applied to this dataset and a two-component model
was selected. Comparing the estimated component membership with the gender yields an ARI $= 0.71$. The contour plot
of the fitted model in Figure~\ref{fig:contor_ais} shows that the fitted model captured the density of
the data fairly well. The Gaussian mixture models and mixtures of generalized hyperbolic distributions are
also applied to the AIS dataset and the summary of the performance are given in Table \ref{tab:ais}. The Gaussian mixture model selects a three component model whereas the mixtures of generalized hyperbolic distribution selected a two component model with slightly higher classification accuracy to the proposed method.\\
\begin{table}
	\caption{Summary of the performances of the proposed model, the Gaussian mixture model, and mixtures of generalized hyperbolic distributions on the AIS datasets.}
	\centering
	\begin{tabular*}{\textwidth}{c@{\extracolsep{\fill}} cc}
		\hline
		&   \textbf{Estimated Groups}   &   \textbf{ARI}    \\\hline
		Proposed Algorithm & $G=2$ & 0.71\\
		GMM &   $G=3$   &   0.69    \\
		MixGHD  &   $G=2$   &   0.77    \\\hline
	\end{tabular*}
	\label{tab:ais}
\end{table}

\section{Conclusion and Discussion}\label{discussion}
In summary, a Dirichlet process mixture of MNIG distributions that
can model skewness and heavy tailed data is proposed. Estimation
of the posterior of the parameters and clustering was done via
Gibbs sampler based on multiple layers of conjugate priors. In our
framework, number of components are treated as a parameter in the
model that moves freely from $1$ to $\infty$ ($N$ in practice),
and are inferred during parameter estimation. This alleviates the
need of choosing a  model selection criteria which can be a
problem. Through simulation studies, we show a near perfect
classification result. Parameter estimates were very close to the
true parameters and our proposed approach provides competitive
results on benchmark datasets while comparing it with state-of-the
art model-based clustering algorithms.\par

Handling of the concentration parameter $\alpha$ for the Dirichlet process prior is still an open problem that need further examination and discussion. Previous work in the context of infinite mixture of elliptical distributions, for example, Gaussian distributions, indicate that smaller $\alpha$ value will encourage the clusters to group together \citep{west1992,escobar1995,BDA3,muller2013}. An alternate approach is to assign a Gamma prior for $\alpha$ and update it in the Gibbs sampling framework to increase flexibility \citep{west1992,escobar1995}. \cite{sylvia2018comparison} in her comparison paper of Dirichlet process mixture models with sparse finite models indicates that one can reduce the prior expectation of number of components by putting a Gamma prior with larger rate parameter to $\alpha$ in the Dirichlet process mixture. We experiment the cases with different values of $\alpha$ by letting $\alpha = 0.5, 1, 2$ and $5$, as well implementing a $Gamma(2,4)$ prior to $\alpha$ (same as \cite{escobar1995} and \cite{sylvia2018comparison}) for both of our simulation studies and the analysis in the fish catch data to test the robustness of clustering results to the choice of $\alpha$. The classification results in both of the simulation studies are shown to be the same. For the analysis of fish catch dataset, with the increasing of values that $\alpha$ takes, in some single Markov chain higher number of components in the model was observed; however, the final models gained when each of the three chains converged and all of them are mixed well (with a potential scale reduction factor smaller than 1.1) tend to always specify the number of clusters as 3. For the cases when algorithm utilizes a higher value of $\alpha$, it is observed that longer Markov chains are required for yielding a stable result. The interesting finding points us to some future directions, such as implementing more advanced sampling technique to improve the efficiency, and inducing some quantification for clustering stability in the Dirichlet process mixture model frameworks.\par

It has been observed that the Dirichlet
process mixture model could overestimate the number of components,
for example, \cite{huelsenbeck2007,onogi2011}, and
\cite{millerharrison2013}. Nevertheless, both
\cite{huelsenbeck2007} and \cite{onogi2011} work with Dirichlet
process mixture of Dirichlet distributions to specify population
structures of genetic allele frequency data, in which, the
overestimated number of components could due to their special
structure, where the choice of concentration parameter to the base
Dirichlet distribution could affect the data allocation too.
Besides, the case discussed in \cite{millerharrison2013} is very
extreme and the conclusion provided is an asymptotic result.
Recently, \cite{yang2019} provide a theoretical study on the lower
bounds on the ratios of posterior probability of number of
components. Some future work will include filling the gap in
understanding influence of choice of $\alpha$ especially in the
mixture of non-elliptical distribution context.



\newpage
\appendix
\section*{Appendix}
\section{Mathematical Details for Posterior Distributions}\label{math_detail}
\subsection{Detail on the posterior updates for the parameters}
Under the Karlis and Santourian (2009) parameterization for MNIG distribution, the mean-variance mixture is of the following structure:
\begin{equation*}
\Xv|u \sim \normal(\muv+u\betav,u\Sigmav),\quad U\sim \IG(1,\gamma)
\end{equation*}
where $\Sigmav$ is not restricted. The density of MNIG distribution is of the form:
\begin{equation*}
f_\Xv(\xv) = \frac{1}{2^{\frac{d-1}{2}}}\left[\frac{\alpha}{\pi q(\xv)}\right]^{\frac{d+1}{2}}\exp\left({p(\xv)}\right)~K_{\frac{d+1}{2}}(\alpha q(\xv))
\end{equation*}
where \[\alpha = \sqrt{\gamma^2 + \betav^\top\Sigmav^{-1}\betav},\quad p(\xv) = \gamma + (\xv - \muv)^\top\Sigmav^{-1}\betav,\quad q(\xv) = \sqrt{1 + (\xv-\muv)^\top \Sigmav^{-1}(\xv-\muv)}\]
and $K_d$ is the modified Bessel function of the third kind of order $d$.\par
The joint probability density of $\xv$ and $u$ is given by
\begin{equation*}
\begin{split}
f(\xv,u) = &f(\xv|u)f(u)\\
= &(2\pi)^{-1/2}|u\Sigmav|^{-1/2}\exp\left\{-\half (\xv - \muv - u\betav)^\top(u\Sigmav)^{-1}(\xv - \muv - u\betav) \right\}\\
&\times \frac{1}{\sqrt{2\pi}}\exp(\gamma)u^{-3/2}\exp\left\{-\half\left(\frac{1}{u}+\gamma^2u\right)\right\}\\
\propto & u^{-\frac{d+3}{2}}|\Sigmav|^{-1/2}\exp\left\{-\half\left(\frac{1 }{u}+\gamma^2u-2\gamma\right) -\half (\xv - \muv - u\betav)^\top(u\Sigmav)^{-1}(\xv - \muv - u\betav)\right\}.
\end{split}
\end{equation*}
The complete data likelihood is of the form:
\begin{equation*}
\begin{split}
L(\theta_g) \propto &\prod_{g=1}^{G}\prod_{i=1}^{N}\left[\pi_g u_{ig}^{-\frac{d+3}{2}}|\Sigmav_g|^{-\half} \exp\left\{-\half\left(u_{ig}^{-1}+\gamma^2u_{ig}-2\gamma_g\right)\right\}\right.\\
&\left.\times \exp\left\{-\half\left(\xv_{i} - \muv_{g} - u_{ig}\betav_{g})^\top(u_{ig}\Sigmav_g)^{-1}(\xv_i - \muv_{g} - u_{ig}\betav_{g}\right)\right\}\right]\\
=&\prod_{g = 1}^{G}\left[\pi_g|\Sigmav_g|^{-\half}\exp\left\{\gamma_g-\betav_g^\top\Sigmav_g^{-1}\muv_g\right\}\right]^{\sumn{z_{ig}}}\times\prod_{g = 1}^{G}\prod_{i=1}^N\left(u_{ig}^{-\frac{d+3}{2}}\right)^{z_{ig}}\\
&\times\prod_{g=1}^{G}\exp\left\{-\half\sumn{z_{ig}u_{ig}^{-1}\xv_i^\top\Sigmav_g^{-1}\xv_i}+\betav_g^\top\Sigmav_g^{-1}\sumn{z_{ig}\xv_i} + \muv_g^\top\Sigmav_g^{-1}\sumn{z_{ig}u_{ig}^{-1}\xv_{i}}\right.\\
&\quad \quad \quad \quad \quad  \left.-\half\left(\betav_g^\top\Sigmav_g^{-1}\betav_g+\gamma_g^2\right)\sumn{z_{ig}u_{ig}} - \half\left(\muv_g^\top\Sigmav_g^{-1}\muv_g+1\right)\sumn{z_{ig}u_{ig}^{-1}}\right\}\\
=&\prod_{g = 1}^{G}\left\{[r(\theta_g)]^{t_{0g}}\cdot\prod_{i = 1}^{N}[h(\xv_i,u_{ig})]^{z_{ig}}\times\exp Tr\left\{\sum_{j = 1}^{5}\phi_{j}(\theta_g)\tv_{jg}(\xv,\uv_{g})\right\}\right\}.\\
\end{split}
\end{equation*}
Then, common conjugate priors and their group specified posterior distributions of each parameter, $\gamma_g, \muv_g, \betav_g, \Sigmav_g$ can be derived in the following.\par
\begin{itemize}
	\item Focusing on $\gamma_g$, the part in the likelihood that contains $\gamma_g$ is as follows
	\begin{equation*}
	\begin{split}
	L(\gamma_g) &\propto \left[\exp\{\gamma_g\}\right]^{\sumn{z_{ig}}}\times\exp\left\{-\half\sumn{z_{ig}u_{ig}}\gamma_g^2\right\}\\ &= \exp\left\{-\half\left(\sumn{z_{ig}u_{ig}}\gamma_g^2-2\sumn{z_{ig}}\gamma_g\right)\right\}=\exp\left\{-\half\left(a_3\gamma_g^2-2a_0\gamma_g\right)\right\}.
	\end{split}
	\end{equation*}
	This is a functional form of normal distribution with mean $\dfrac{a_0}{a_3}$ and variance $\dfrac{1}{a_3}$, truncated at 0 because we want $\gamma_g$ to be positive. Therefore, a conjugate truncated normal prior is assigned to $\gamma_g$, i.e.
	\begin{equation*}
	\gamma_g \sim \normal \left(\frac{a_0^{(0)}}{a_3^{(0)}},\frac{1}{a_3^{(0)}}\right)\cdot\1v\left(\gamma_g > 0\right);
	\end{equation*}
	and the resulting posterior is truncated normal as well:
	\begin{equation*}
	\gamma_g \sim \normal\left(\dfrac{a_{0,g}}{a_{3,g}},\dfrac{1}{a_{3,g}}\right)\cdot\1v\left(\gamma_g > 0\right).
	\end{equation*}
	\item For the precision matrix $\Sigmav_g^{-1}$, denoted as $\Tv_g$, in the following derivation, the part of likelihood which contains $\Tv_g=\Sigmav_g^{-1}$ yields the following:
	\begin{equation*}
	\begin{split}
	L(\Tv_g) &\propto |\Tv_g|^{\sumn{z_{ig}}/2}\cdot\exp\left\{-\half Tr\left[\sumn{z_{ig}u_{ig}^{-1}\left(\xv_{i}-\muv_g-u_{ig}\betav_g\right)\left(\xv_{i}-\muv_g-u_{ig}\betav_g\right)^\top}\Tv_g\right]\right\}\\
	&=|\Tv_g|^{a_0/2}\exp\left\{-\half Tr\left(\Vv_0^{-1}\Tv_g\right)\right\};
	\end{split}
	\end{equation*}
	which is a functional form of the Wishart distribution.\par
	Hence, a conjugate $Wishart\left(a_0^{(0)},{\av_5^{(0)}}^{-1}\right)$ prior is given to $\Tv_g$. When computing the posterior, we take $(\muv_g,\betav_g)$ into consideration as well because they carry information about $\Tv_g$ and in addition, they contribute to likelihood together with $\Tv_g$; therefore the resulting posterior is conditional on $(\muv_g,\betav_g)$, and is of the form
	\begin{equation*}
	\Tv_g|(\muv_g,\betav_g) \sim Wishart(a_{0,g},\Vv_{0,g}).
	\end{equation*}
	The derivation of the posterior parameter $\Vv_{0,g}$ is presented at the end of this section.
	\item Look at the pair of $(\muv_g,\betav_g)$, whose related part in the likelihood is conditional on $\Tv_g = \Sigmav_g^{-1}$ and is a functional form of correlated Gaussian distribution.
	\begin{equation*}
	\begin{split}
	L(\muv_g,\betav_g|\Tv_g) & \propto \exp\left\{\sumn{z_{ig}\left(-\betav_{g}^\top \Tv_g \muv_{g}\right)}\right.\\
	&\quad \quad \quad \quad \quad \left.-\half \left[\sumn{z_{ig}u_{ig}^{-1}\left(\xv_{i}-\muv_g-u_{ig}\betav_g\right)^\top \Tv_g\left(\xv_{i}-\muv_g-u_{ig}\betav_g\right)}\right]\right\}\\
	&\propto \exp\left\{\betav_{g}^\top \Tv_g\muv_{g}\sumn{z_{ig}}+\betav_g^\top \Tv_g\sumn{z_{ig}\xv_i} + \muv_g^\top \Tv_g\sumn{z_{ig}u_{ig}^{-1}\xv_{i}}\right.\\
	&\quad \quad \quad \quad \quad  \left.-\half\left(\betav_g^\top \Tv_g\betav_g\right)\sumn{z_{ig}u_{ig}} - \half\left(\muv_g^\top \Tv_g\muv_g\right)\sumn{z_{ig}u_{ig}^{-1}}\right\}.
	\end{split}
	\end{equation*}
	Therefore, a multivariate normal distribution conditional on the precision matrix is assigned to $(\muv_g,\betav_g)$ such that
	\begin{equation*}
	\left.\begin{pmatrix} \muv_g\\\betav_g\end{pmatrix}\right\vert \Tv_g\sim\normal\left[\begin{pmatrix} \muv_0^{(0)}\\ \betav_0^{(0)}\end{pmatrix},\begin{pmatrix}
	\tau_\mu^{(0)} \Tv_g & \tau_{\mu\beta}^{(0)} \Tv_g\\
	\tau_{\mu\beta}^{(0)} \Tv_g & \tau_\beta^{(0)} \Tv_g
	\end{pmatrix}\right];
	\end{equation*}
	and hence the posterior is then
	\begin{equation*}
	\left.\begin{pmatrix} \muv_g\\\betav_g\end{pmatrix}\right\vert \Tv_g\sim\normal\left[\begin{pmatrix} \muv_{0,g}\\ \betav_{0,g}\end{pmatrix},\begin{pmatrix}
	\tau_{\mu,g} \Tv_g & \tau_{\mu\beta,g} \Tv_g\\
	\tau_{\mu\beta,g} \Tv_g & \tau_{\beta,g} \Tv_g
	\end{pmatrix}\right].
	\end{equation*}
	where \begin{align*}
	\muv_0^{(0)} &= \frac{a_{3}^{(0)}\av_{2}^{(0)} - a_{0}^{(0)}\av_{1}^{(0)}}{a_{3}^{(0)}a_{4}^{(0)}-{a_{0}^{(0)}}^2}, &\muv_{0,g} &= \frac{a_{3,g}\av_{2,g} - a_{0,g}\av_{1,g}}{a_{3,g}a_{4,g}-{a_{0,g}}^2};\\
	\betav_0^{(0)} &= \frac{a_{4}^{(0)}\av_{1}^{(0)} - a_{0}^{(0)}\av_{2}^{(0)}}{a_{3}^{(0)}a_{4}^{(0)}-{a_{0}^{(0)}}^2}, &\betav_{0,g} &= \frac{a_{4,g}\av_{1,g} - a_{0,g}\av_{2,g}}{a_{3,g}a_{4,g}-{a_{0,g}}^2};\\
	\tau_\mu^{(0)} &= a_{4}^{(0)}, &\tau_{\mu,g} &= a_{4,g};\\
	\tau_\beta^{(0)} &= a_{3}^{(0)}, &\tau_{\beta,g} &= a_{3,g};\\
	\tau_{\mu\beta}^{(0)} &= a_{0}^{(0)}, &\tau_{\mu\beta,g} &= a_{0,g}.
	\end{align*}
	Derivation of the forms of $\muv_{0,g},\betav_{0,g},\tau_{\mu,g},\tau_{\beta,g},$ and $\tau_{\mu\beta,g}$ is shown as the follows.
	\item Now, recall that the prior of $\Tv_g = \Sigmav_g^{-1}$ is $\Tv_g \sim Wishart\left(a_0^{(0)},{\av_5^{(0)}}^{-1}\right)$; also, conditional on $\Tv_g$, $(\muv_g,\betav_g)$ follows a multivariate normal distribution jointly. Consider these two prior densities jointly, we derive the updated form of $\Vv_{0,g},\muv_{0,g},\betav_{0,g},\tau_{\mu,g},\tau_{\beta,g},$ and $\tau_{\mu\beta,g}$.\par
	The joint prior density of $\muv_{g},\betav_{g}, \Tv_g$ is as following:
	\begin{equation*}
	\begin{split}
	p(\muv_g,\betav_g,\Tv_g) \propto \begin{vmatrix}
	\tau_\mu^{(0)} \Tv_g & \tau_{\mu\beta}^{(0)} \Tv_g\\
	\tau_{\mu\beta}^{(0)} \Tv_g & \tau_\beta^{(0)} \Tv_g
	\end{vmatrix}^{-\half}&\exp\left\{-\half\begin{pmatrix} \muv_g-\muv_0^{(0)}\\\betav_g-\betav_0^{(0)}\end{pmatrix}^\top\begin{pmatrix}
	\tau_\mu^{(0)} \Tv_g & \tau_{\mu\beta}^{(0)} \Tv_g\\
	\tau_{\mu\beta}^{(0)} \Tv_g & \tau_\beta^{(0)} \Tv_g
	\end{pmatrix}\begin{pmatrix} \muv_g-\muv_0^{(0)}\\\betav_g-\betav_0^{(0)}\end{pmatrix}\right\}\\
	\times \left\vert \Tv_g\right\vert^{\frac{(a_0^{(0)}+d+1)-d-1}{2}}&\exp\left\{-\half Tr\left(\av_5^{(0)}\Tv_g\right)\right\}
	\end{split}
	\end{equation*}
	Look at the part inside the $\exp$ function only, we have a form of the following:
	\begin{equation*}
	\begin{split}
	&Tr\left(\av_5^{(0)}\Tv_g\right)+(\muv_g-\muv_0^{(0)})^\top\tau_\mu^{(0)}\Tv_g(\muv_g-\muv_0^{(0)})+(\betav_g-\betav_0^{(0)})^\top\tau_{\mu\beta}^{(0)}\Tv_g(\muv_g-\muv_0^{(0)})\\
	&\quad\quad \quad \quad \quad+(\muv_g-\muv_0^{(0)})^\top\tau_{\mu\beta}^{(0)}\Tv_g(\betav_g-\betav_0^{(0)})+(\betav_g-\betav_0^{(0)})^\top\tau_\beta^{(0)}\Tv_g(\betav_g-\betav_0^{(0)})\\
	=& Tr\left(\av_5^{(0)}\Tv_g\right)+\muv_g^\top\tau_\mu^{(0)}\Tv_g\muv_g-2\muv_g^\top\tau_\mu^{(0)}\Tv_g\muv_0^{(0)}-2\muv_g^\top\tau_{\mu\beta}^{(0)}\Tv_g\betav_0^{(0)}\\
	&\quad\quad \quad \quad \quad +\betav_g^\top\tau_\beta^{(0)}\Tv_g\betav_g-2\betav_g^\top\tau_\beta^{(0)}\Tv_g\betav_0^{(0)}-2\betav_g^\top\tau_{\mu\beta}^{(0)}\Tv_g\muv_0^{(0)}\\
	&\quad\quad \quad \quad \quad + 2\betav_g^\top\tau_{\mu\beta}^{(0)}\Tv_g\muv_g+{\muv_0^{(0)}}^\top\tau_\mu^{(0)}\Tv_g\muv_0^{(0)}+2{\betav_0^{(0)}}^\top\tau_{\mu\beta}^{(0)}\Tv_g\muv_0^{(0)}+{\betav_0^{(0)}}^\top\tau_\beta^{(0)}\Tv_g\betav_0^{(0)}.
	\end{split}
	\end{equation*}
	Now multiply the joint prior density with the likelihood that contains $\muv_g,\betav_g$, and $\Tv_g$, we get the posterior. Only look at the part inside the $\exp$ again, we have the following:
	\begin{equation*}
	\begin{split}
	&Tr\left(\av_5^{(0)}\Tv_g\right)+(\muv_g-\muv_0^{(0)})^\top\tau_\mu^{(0)}\Tv_g(\muv_g-\muv_0^{(0)})+(\betav_g-\betav_0^{(0)})^\top\tau_{\mu\beta}^{(0)}\Tv_g(\muv_g-\muv_0^{(0)})\\
	&\quad\quad \quad \quad \quad +(\muv_g-\muv_0^{(0)})^\top\tau_{\mu\beta}^{(0)}\Tv_g(\betav_g-\betav_0^{(0)})+(\betav_g-\betav_0^{(0)})^\top\tau_\beta^{(0)}\Tv_g(\betav_g-\betav_0^{(0)})\\
	&\quad\quad \quad \quad \quad +\sumn{z_{ig}u_{ig}^{-1}}\left(\xv_i-\muv_g-u_{ig}\betav_g\right)^\top \Tv_g \left(\xv_i-\muv_g-u_{ig}\betav_g\right)\\
	=& Tr\left(\av_5^{(0)}\Tv_g\right)+\muv_g^\top\tau_\mu^{(0)}\Tv_g\muv_g-2\muv_g^\top\tau_\mu^{(0)}\Tv_g\muv_0^{(0)} - 2\muv_g^\top\tau_{\mu\beta}^{(0)}\Tv_g\betav_0^{(0)}\\
	&\quad\quad \quad \quad \quad - 2\sumn{z_{ig}u_{ig}^{-1}}\xv_{i}\Tv_g\muv_g + \muv_g^\top\sumn{z_{ig}u_{ig}^{-1}}\Tv_g\muv_g\\
	&\quad\quad \quad \quad \quad + \betav_{g}^\top\tau_\beta^{(0)}\Tv_g\betav_{g} - 2\betav_{g}^\top\tau_\beta^{(0)}\Tv_g\betav_0^{(0)} - 2\betav_g^\top\tau_{\mu\beta}^{(0)}\Tv_g\muv_0^{(0)}\\
	&\quad\quad \quad \quad \quad - 2\sumn{z_{ig}\xv_i^\top}\Tv_g\betav_g + \betav_g^\top\sumn{z_{ig}u_{ig}}\Tv_g\betav_g\\
	&\quad\quad \quad \quad \quad+ 2\muv_g^\top\tau_{\mu\beta}^{(0)}\Tv_g\betav_g  + 2\muv_g^\top\sumn{z_{ig}}\Tv_g\betav_g\\
	&\quad\quad \quad \quad \quad + 2{\muv_0^{(0)}}^\top\tau_{\mu\beta}^{(0)}\Tv_g\betav_0^{(0)} + {\muv_0^{(0)}}^\top\tau_\mu^{(0)}\Tv_g\muv_0^{(0)} + {\betav_0^{(0)}}^\top\tau_\beta^{(0)}\Tv_g\betav_0^{(0)}+\sumn{z_{ig}u_{ig}^{-1}}\xv_{i}^\top \Tv_g \xv_{i}.
	\end{split}
	\end{equation*}
	The exponential part inside the posterior density should have the same functional form as the prior, i.e., it is expected to have the form of
	\begin{equation*}
	\begin{split}
	&Tr\left(\Vv_{0,g}^{-1}\Tv_g\right)+\muv_g^\top\tau_\mu \Tv_g\muv_g-2\muv_g^\top\tau_\mu \Tv_g\muv_{0,g}-2\muv_g^\top\tau_{\mu\beta} \Tv_g\betav_{0,g}\\
	&\quad\quad \quad \quad \quad +\betav_g^\top\tau_\beta \Tv_g\betav_g-2\betav_g^\top\tau_\beta \Tv_g\betav_{0,g}-2\betav_g^\top\tau_{\mu\beta} \Tv_g\muv_{0,g}\\
	&\quad\quad \quad \quad \quad + 2\betav_g^\top\tau_{\mu\beta} \Tv_g\muv_g+{\muv_{0,g}}^\top\tau_\mu \Tv_g\muv_{0,g}+2\betav_{0,g}^\top\tau_{\mu\beta} \Tv_g\muv_{0,g}+{\betav_{0,g}}^\top\tau_\beta \Tv_g\betav_{0,g}.
	\end{split}
	\end{equation*}
	By comparing the previous two terms, we derive the following:
	\begin{equation*}
	\begin{split}
	\tau_{\mu,g} &= \tau_\mu^{(0)}+\sumn{z_{ig}u_{ig}^{-1}} = a_4^{(0)}+\sumn{z_{ig}u_{ig}^{-1}} = a_{4,g};\\
	\tau_{\beta,g} &= \tau_\beta^{(0)}+\sumn{z_{ig}u_{ig}} = a_3^{(0)}+\sumn{z_{ig}u_{ig}} = a_{3,g};\\
	\tau_{\mu\beta,g} &=\tau_{\mu\beta}^{(0)}+\sumn{z_{ig}} = a_0^{(0)}+\sumn{z_{ig}} = a_{0,g}.
	\end{split}
	\end{equation*}
	Besides, we have
	\begin{equation*}
	\begin{split}
	&\left\{\begin{split}
	\tau_{\mu,g}\Tv_g\muv_{0,g} + \tau_{\mu\beta,g}\Tv_g\betav_{0,g} &= \tau_\mu^{(0)}\Tv_g\muv_0^{(0)} + \tau_{\mu\beta}^{(0)}\Tv_g\betav_0^{(0)}+\Tv_g\sumn{z_{ig}u_{ig}^{-1}\xv_{i}}\\
	\tau_{\mu\beta,g}\Tv_g\muv_{0,g} + \tau_{\beta,g}\Tv_g\betav_{0,g} &= \tau_\beta^{(0)}\Tv_g\betav_0^{(0)} + \tau_{\mu\beta}^{(0)}\Tv_g\muv_0^{(0)}+\Tv_g\sumn{z_{ig}\xv_{i}}
	\end{split}\right.\\
	\Rightarrow &\left\{\begin{split}
	a_{4,g}\muv_{0,g} + a_{0,g}\betav_{0,g} &= a_4^{(0)}\muv_0^{(0)} + a_0^{(0)}\betav_0^{(0)}+\sumn{z_{ig}u_{ig}^{-1}\xv_{i}}\\
	a_{0,g}\muv_{0,g} + a_{3,g}\betav_{0,g} &= a_3^{(0)}\betav_0^{(0)} + a_0^{(0)}\muv_0^{(0)}+\sumn{z_{ig}\xv_{i}}
	\end{split}\right.
	\end{split};
	\end{equation*}
	plus given that \begin{equation*}
	\muv_0^{(0)} = \frac{a_{3}^{(0)}\av_{2}^{(0)} - a_{0}^{(0)}\av_{1}^{(0)}}{a_{3}^{(0)}a_{4}^{(0)}-{a_{0}^{(0)}}^2}, \quad \quad
	\betav_0^{(0)} = \frac{a_{4}^{(0)}\av_{1}^{(0)} - a_{0}^{(0)}\av_{2}^{(0)}}{a_{3}^{(0)}a_{4}^{(0)}-{a_{0}^{(0)}}^2},
	\end{equation*}
	the solution to $\muv_{0,g}$ and $\betav_{0,g}$ is
	\begin{equation*}
	\muv_{0,g} = \frac{a_{3,g}\av_{2,g} - a_{0,g}\av_{1,g}}{a_{3,g}a_{4,g}-{a_{0,g}}^2}, \quad \quad \betav_{0,g} = \frac{a_{4,g}\av_{1,g} - a_{0,g}\av_{2,g}}{a_{3,g}a_{4,g}-{a_{0,g}}^2}.
	\end{equation*}
	Last, we have
	\begin{equation*}
	\begin{split}
	&Tr\left(\Vv_{0,g}^{-1}\Tv_g\right)+\muv_g^\top\tau_\mu \Tv_g\muv_g +\betav_g^\top\tau_\beta \Tv_g\betav_g + 2\betav_g^\top\tau_{\mu\beta} \Tv_g\muv_g\\
	=&Tr\left(\av_5^{(0)}\Tv_g\right)+2{\muv_0^{(0)}}^\top\tau_{\mu\beta}^{(0)}\Tv_g\betav_0^{(0)} + {\muv_0^{(0)}}^\top\tau_\mu^{(0)}\Tv_g\muv_0^{(0)} + {\betav_0^{(0)}}^\top\tau_\beta^{(0)}\Tv_g\betav_0^{(0)}+\sumn{z_{ig}u_{ig}^{-1}}\xv_{i}^\top \Tv_g \xv_{i};
	\end{split}
	\end{equation*}
	from which $\Vv_{0,g}^{-1}$ is shown to have the form as below:
	\begin{equation*}
	\begin{split}
	\Vv_{0,g}^{-1} = &\av_5^{(0)} + \sumn{z_{ig}u_{ig}^{-1}\xv_{i}\xv_{i}^\top}\\
	&+\muv_0^{(0)}\tau_{\mu}^{(0)}{\muv_0^{(0)}}^\top+\muv_0^{(0)}\tau_{\mu\beta}^{(0)}{\betav_0^{(0)}}^\top+\betav_0^{(0)}\tau_{\mu\beta}^{(0)}{\muv_0^{(0)}}^\top+\betav_0^{(0)}\tau_{\beta}^{(0)}{\betav_0^{(0)}}^\top\\
	&-\left(\muv_{0,g}\tau_{\mu,g}\muv_{0,g}^\top+\muv_{0,g}\tau_{\mu\beta,g}\betav_{0,g}^\top+\betav_{0,g}\tau_{\mu\beta,g}\muv_{0,g}^\top+\betav_{0,g}\tau_{\beta,g}\betav_{0,g}^\top\right)\\
	=&\av_{5,g}+\muv_0^{(0)}\tau_{\mu}^{(0)}{\muv_0^{(0)}}^\top+\muv_0^{(0)}\tau_{\mu\beta}^{(0)}{\betav_0^{(0)}}^\top+\betav_0^{(0)}\tau_{\mu\beta}^{(0)}{\muv_0^{(0)}}^\top+\betav_0^{(0)}\tau_{\beta}^{(0)}{\betav_0^{(0)}}^\top\\
	&-\left(\muv_{0,g}\tau_{\mu,g}\muv_{0,g}^\top+\muv_{0,g}\tau_{\mu\beta,g}\betav_{0,g}^\top+\betav_{0,g}\tau_{\mu\beta,g}\muv_{0,g}^\top+\betav_{0,g}\tau_{\beta,g}\betav_{0,g}^\top\right)\\
	\end{split}
	\end{equation*}
\end{itemize}
\subsection{Detail on posterior updates for hyperparameters}
\subsubsection{Derivation of the posteriors}
There are six hyperparameters, $a_0^{(0)}, \av_{1}^{(0)}, \av_{2}^{(0)}, a_3^{(0)}, a_4^{(0)},$ and $\av_5^{(0)}$, common
for all groups, that defines the prior distributions of the parameters. Another layer of prior distributions can be added on
these hyperparameters for additional flexibility. Again, common conjugate prior distributions are assigned to these
hyperparameters, which yield posteriors that depends on the group-specified observations. Details of the derivation are given below.\par
Again, recall that the complete data likelihood can be written into a form that comes from the exponential family:
\begin{equation*}
L(\theta_g)\propto \prod_{g = 1}^{G}\left\{[r(\theta_g)]^{t_{0g}}\cdot\prod_{i = 1}^{N}[h(\xv_i,u_{ig})]^{z_{ig}}\times\exp Tr\left\{\sum_{j = 1}^{5}\phi_{j}(\theta_g)\tv_{jg}(\xv,\uv_{g})\right\}\right\}
\end{equation*}
where \begin{align*}
r(\thetav_g) &= \pi_g|\Sigmav_g|^{-\half}\exp\left\{\gamma_g - \muv_g^\top\Sigmav_g^{-1}\betav_g\right\}, &t_{0g} &= \sumn{z_{ig}};\\
\phi_{1}(\thetav_g) &= \Sigmav_g^{-1}\betav_g, &\tv_{1g} (\xv,\uv_{g})&= \sumn z_{ig}\xv_i^{\top};\\
\phi_{2}(\thetav_g) &= \Sigmav_g^{-1}\muv_g, &\tv_{2g} (\xv,\uv_{g})&= \sumn z_{ig}u_{ig}^{-1}\xv_{i}^\top;\\
\phi_{3}(\thetav_g) &= -\half (\Sigmav_g^{-1}\betav_g\betav_g^\top + \frac{\gamma_g^2}{d}\Id_{d}), &t_{3g}(\xv,\uv_{g}) &= \sumn z_{ig}u_{ig};\\
\phi_{4}(\thetav_g) &= -\half (\Sigmav_g^{-1}\muv_g\muv_g^\top + \frac{1}{d}\Id_{d}), &t_{4g} (\xv,\uv_{g})&= \sumn z_{ig}u_{ig}^{-1};\\
\phi_{5}(\thetav_g) &= -\half \Sigmav_g^{-1}, &\tv_{5g} (\xv,\uv_{g})&= \sumn z_{ig} u_{ig}^{-1}\xv_i\xv_i^\top.
\end{align*}
\begin{itemize}
	\item $a_0^{(0)}$ is associated with $t_{0g}$, who only relates to $r(\thetav_g)$ in the complete data likelihood, with a functional form of the density from an exponential distribution. Therefore, an exponential prior with rate parameter $b_0$ is assigned to $a_0^{(0)}$:
	\begin{equation*}
	a_0^{(0)} \sim \text{Exp}(b_0),\quad  p\left(a_0^{(0)}\right) = b_0\exp\left(-b_0a_0^{(0)}\right);
	\end{equation*}
	hence the posterior is an exponential distribution as well with rate parameter  $b_0-\sum_{g=1}^{G}\log(\pi_g)-\sum_{g=1}^{G}\log\left(|\Sigmav_g|^{-\half}\right)-\sum_{g=1}^{G}\gamma_g+\sum_{g=1}^{G}\muv_g^\top\Sigmav_g^{-1}\betav_g$:
	\begin{equation*}
	\begin{split}
	p\left(a_0^{(0)}|r(\theta_1),\dots r(\theta_G)\right)&\propto \exp\left\{-b_0a_0^{(0)}\right\}\prod_{g=1}^{G}\exp\left\{\left[r(\theta_g)\right]^{t_{0g}}\right\}\\
	&=\exp\left\{-b_0a_0^{(0)}\right\}\exp\left\{\sum_{g=1}^{G}\left[\log(\pi_g)+\log\left(|\Sigmav_g|^{-\half}\right)+\gamma_g-\muv_g^\top\Sigmav_g^{-1}\betav_g\right]t_{0g}\right\}\\
	&=\exp\left\{-\left[b_0-\sum_{g=1}^{G}\log(\pi_g)-\sum_{g=1}^{G}\log\left(|\Sigmav_g|^{-\half}\right)\right.\right.\\
	&\quad \quad \quad \quad \quad \quad \quad \quad  \left.\left.-\sum_{g=1}^{G}\gamma_g+\sum_{g=1}^{G}\muv_g^\top\Sigmav_g^{-1}\betav_g\right]a_0^{(0)}\right\}
	\end{split}
	\end{equation*}
	\item $\av_{1}^{(0)}$ is associated with $\tv_{1g}$, whose functional form in the likelihood is a multivariate Gaussian distribution and relate to $\phi_{1}(\theta_g)$ only. A multivariate Gaussian prior is then assigned to $\av_{1}^{(0)}$; it yields a multivariate Gaussian posterior:
	\begin{equation*}
	\av_{1}^{(0)} \sim \normal(\cv_1,B_1),\quad p\left(\av_{1}^{(0)}\right)\propto \exp\left\{-\half\left(\av_{1}^{(0)}-\cv_1\right)^\top B_1^{-1}\left(\av_{1}^{(0)}-\cv_1\right)\right\};
	\end{equation*}
	\begin{equation*}
	\begin{split}
	p\left(\av_{1}^{(0)}|\phi_{1}(\theta_1),\dots,\phi_{1}(\theta_G)\right)&\propto \exp\left\{-\half\left(\av_{1}^{(0)}-\cv_1\right)^\top B_1^{-1}\left(\av_{1}^{(0)}-\cv_1\right)\right\} \prod_{g=1}^{G}\exp Tr\left\{\phi_{1g}\tv_{1g}\right\}\\
	&=\exp\left\{-\half\left(\av_{1}^{(0)}-\cv_1\right)^\top B_1^{-1}\left(\av_{1}^{(0)}-\cv_1\right)+\sum_{g=1}^{G}\betav_g^\top\Sigmav_g^{-1} \tv_{1g}\right\}\\
	&=\exp\left\{-\half\left[{\av_{1}^{(0)}}^\top B_1^{-1}\av_{1}^{(0)}\right.\right.\\
	&\quad \quad \quad \quad \quad \quad
	\left.\left.-2\left(\cv_1^\top B_1^{-1}+\sum_{g=1}^{G}\betav_g^\top\Sigmav_g^{-1}\right)\av_{1}^{(0)}+\cv_1B_1^{-1}\cv_1 \right]\right\}\\
	&\propto \exp\left\{-\half\left[\av_{1}^{(0)} - \left(\cv_1+\sum_{g=1}^{G}\betav_g^\top\Sigmav_g^{-1}B_1\right)B_1^{-1}\right]^\top\right. \\
	&\quad \quad \quad \quad \quad \left.B_1^{-1}\left[\av_{1}^{(0)} - \left(\cv_1+\sum_{g=1}^{G}\betav_g^\top\Sigmav_g^{-1}B_1\right)B_1^{-1}\right]\right\}.
	\end{split}
	\end{equation*}
	The mean and covariance matrix of the posterior is $\cv_1+\sum_{g=1}^{G}\betav_g^\top\Sigmav_g^{-1}B_1$ and $B_1$ respectively.
	\item Similar to $\av_{1}^{(0)}$, $\av_{2}^{(0)}$ is associated with $\tv_{2g}$ who also possess a functional term of multivariate Gaussian that relate to the part including $\phi_{2}(\theta_g)$ only in the likelihood. A multivariate Gaussian prior is assigned to $\av_{2}^{(0)}$ and results in a multivariate Gaussian posterior with mean $\cv_2+\sum_{g=1}^{G}\muv_g^\top\Sigmav_g^{-1}B_2$ and covariance $B_2$.
	\begin{equation*}
	\av_{2}^{(0)} \sim \normal(\cv_2,B_2),\quad p\left(\av_{2}^{(0)}\right)\propto \exp\left\{-\half\left(\av_{2}^{(0)}-\cv_2\right)^\top B_2^{-1}\left(\av_{2}^{(0)}-\cv_2\right)\right\};
	\end{equation*}
	\begin{equation*}
	\begin{split}
	p\left(\av_{2}^{(0)}|\phi_{2}(\theta_1),\dots,\phi_{2}(\theta_G)\right) &\propto \exp\left\{-\half\left(\av_{2}^{(0)}-\cv_2\right)^\top B_2^{-1}\left(\av_{2}^{(0)}-\cv_2\right)\right\} \prod_{g=1}^{G}\exp Tr\left\{\phi_{2g}\tv_{2g}\right\}\\
	&=\exp\left\{-\half\left(\av_{2}^{(0)}-\cv_2\right)^\top B_2^{-1}\left(\av_{2}^{(0)}-\cv_2\right)+\sum_{g=1}^{G}\muv_g^\top\Sigmav_g^{-1} \tv_{2g}\right\}\\
	&=\exp\left\{-\half\left[{\av_{2}^{(0)}}^\top B_2^{-1}\av_{2}^{(0)}\right.\right.\\
	&\quad \quad \quad \quad \quad \quad
	\left.\left.-2\left(\cv_2^\top B_2^{-1}+\sum_{g=1}^{G}\muv_g^\top\Sigmav_g^{-1}\right)\av_{2}^{(0)}+\cv_2B_2^{-1}\cv_2 \right]\right\}\\
	&\propto \exp\left\{-\half\left[\av_{2}^{(0)} - \left(\cv_2+\sum_{g=1}^{G}\muv_g^\top\Sigmav_g^{-1}B_2\right)B_2^{-1}\right]^\top \right.\\
	&\quad \quad \quad \quad \quad \left. B_2^{-1}\left[\av_{2}^{(0)} - \left(\cv_2+\sum_{g=1}^{G}\muv_g^\top\Sigmav_g^{-1}B_2\right)B_2^{-1}\right]\right\}.
	\end{split}
	\end{equation*}
	\item $a_3^{(0)}$ is associated with $t_{3g}$, which only related to $\phi_{3}(\theta_g)$ in the likelihood and has a functional form of an exponential distribution; hence $a_3^{(0)}$ is assigned an exponential prior with a resulting posterior being exponential too.
	\begin{equation*}
	a_3^{(0)} \sim \text{Exp}(b_3),\quad p\left(a_3^{(0)}\right) = b_3\exp\left(-b_3a_3^{(0)}\right);
	\end{equation*}
	\begin{equation*}
	\begin{split}
	p\left(a_3^{(0)}|\phi_{3}(\theta_1),\dots\phi_{3}(\theta_G)\right) &\propto \exp\left\{-b_3a_3^{(0)}\right\}\prod_{g=1}^{G}\exp\left\{\phi_{3g}t_{3g}\right\}\\
	&=\exp\left\{-\left[b_3 - \half\sum_{g=1}^{G}\left(\beta_g^\top\Sigmav_g^{-1}\beta_g+\gamma_g^2\right)\right]a_3^{(0)}\right\};
	\end{split}
	\end{equation*}
	where in the posterior distribution, the rate parameter is $b_3 - \half\sum_{g=1}^{G}\left(\betav_g^\top\Sigmav_g^{-1}\betav_g+\gamma_g^2\right)$.
	\item Similar to $a_3^{(0)}$, $a_4^{(0)}$ is assigned an exponential prior:
	\begin{equation*}
	a_4^{(0)} \sim \text{Exp}(b_4),\quad p\left(a_4^{(0)}\right) = b_4\exp\left(-b_4a_4^{(0)}\right);
	\end{equation*}
	\begin{equation*}
	\begin{split}
	p\left(a_4^{(0)}|\phi_{4}(\theta_1),\dots,\phi_{4}(\theta_G)\right) &\propto \exp\left\{-b_4a_4^{(0)}\right\}\prod_{g=1}^{G}\exp\left\{\phi_{4g}t_{4g}\right\}\\
	&=\exp\left\{-\left[b_4 - \half\sum_{g=1}^{G}\left(\muv_g^\top\Sigmav_g^{-1}\muv_g+1\right)\right]a_4^{(0)}\right\}.
	\end{split}
	\end{equation*}
	It indicates that the posterior is exponential distribution with rate $b_4 - \half\sum_{g=1}^{G}\left(\muv_g^\top\Sigmav_g^{-1}\muv_g+1\right)$.
	\item $\av_5^{(0)}$ is associated with $\Sigmav_g^{-1}$. A Wishart prior is assigned to $\av_5^{(0)}$; the resulting posterior is also a Wishart distribution.
	\begin{equation*}
	\av_5^{(0)} \sim Wishart(\nu_0,\Lambdav_0),\quad p\left(\av_5^{(0)}\right)\propto \left\vert \av_5^{(0)}\right\vert^{\frac{\nu_0-d-1}{2}}\exp\left\{-\half Tr \left(\Lambdav_0^{-1}\av_5^{(0)}\right)\right\}
	\end{equation*}
	The priors of $\Sigmav_g^{-1}$ perform as the likelihood in the derivation of the posterior of $\av_5^{(0)}$:
	\begin{equation*}
	\begin{split}
	p\left(\left.\av_5^{(0)}\right\vert \Sigmav_1^{-1},\dots,\Sigmav_G^{-1}\right) &\propto\left\vert\av_5^{(0)}\right\vert^{\frac{\nu_0-d-1}{2}}\exp\left\{-\half Tr \left(\Lambdav_0^{-1}\av_5^{(0)}\right)\right\}\times \prod_{g=1}^{G}\left\vert\av_5^{(0)}\right\vert^{\frac{a_0^{(0)}}{2}}\exp\left\{-\half Tr \left(\av_5^{(0)}\Sigmav_g^{-1}\right)\right\}\\
	&= \left\vert\av_5^{(0)}\right\vert^{\frac{\nu_0+G\times a_0^{(0)}-d-1}{2}}\exp\left\{-\half Tr \left[\left(\Lambdav_0^{-1}+\sum_{g=1}^{G}\Sigmav_g^{-1}\right)\av_5^{(0)}\right]\right\}
	\end{split}
	\end{equation*}
	Therefore, the posterior of $\av_5^{(0)}$ is $Wishart\left[\nu_0+G\times a_0^{(0)},\left(\Lambdav_0^{-1}+\sum_{g=1}^{G}\Sigmav_g^{-1}\right)^{-1}\right]$
\end{itemize}
\subsubsection{Choice of third layer hyperparameters}
The third layer hyperparameters $b_0,c_1,B_1,c_2,B_2,b_3,b_4,\nu_0$ and $\Lambdav_0$ are chosen such that if a sample is drawn from the posteriors of the hyperparameters $a_0^{(0)},\av_{1}^{(0)},\av_{2}^{(0)},a_3^{(0)},a_4^{(0)}$ and $\av_5^{(0)}$, it is expected to close to the associated term of $t_{0g},\tv_{1g}, \tv_{2g}, t_{3g}, t_{4g}$ when $g=G=1$ and the sample covariance matrix $\Sigmav_{\xv}$, respectively. Therefore, we have the following:\\
When $G=1$, $z_1 = z_2 = \cdots = z_N$ hence $z_{ig}=1$ when $i=g$ and $z_{ig}= 0$ for $\forall i\neq g$.
\begin{align*}
1/b_0 &= \E\left(a_0^{(0)}\right) \sim \sumn{z_{ig}}=N, &\to \quad \quad b_0 &= 1/N;\\
1/b_3 &= \E\left(a_3^{(0)}\right) \sim \sumn{z_{ig}u_{ig}}, &\to \quad \quad b_3 &= 1/\sumn{u_{i}};\\
1/b_4 &= \E\left(a_4^{(0)}\right) \sim \sumn{z_{ig}u_{ig}^{-1}}, &\to \quad \quad  b_4 &= 1/\sumn{u_{i}^{-1}};\\
\cv_1 &= \E\left(\av_{1}^{(0)}\right) \sim \sumn{z_{ig}\xv_{i}} &\to \quad\quad  \cv_1 &= \sumn{\xv_{i}}, B_1 = \Sigmav_{\xv};\\
\cv_2 &= \E\left(\av_{2}^{(0)}\right) \sim \sumn{z_{ig}u_{ig}^{-1}\xv_{i}} &\to \quad \quad \cv_2 &= \sumn{u_{i}^{-1}\xv_{i}}, B_2 = \Sigmav_{u^{-1}\xv};\\
\nu_0\Lambdav_0 &= \E\left(\av_5^{(0)}\right) \sim \Sigmav_{\xv} &\to \quad \quad  \Lambdav_0 &= \Sigmav_{\xv}/\nu_0, \nu_0 = d+1.
\end{align*}
\newpage
\section{Credible Intervals for Simulation Study 1}\label{supp_fig}
\begin{figure*}[!h]
	\centering
	\includegraphics[width=.7\linewidth]{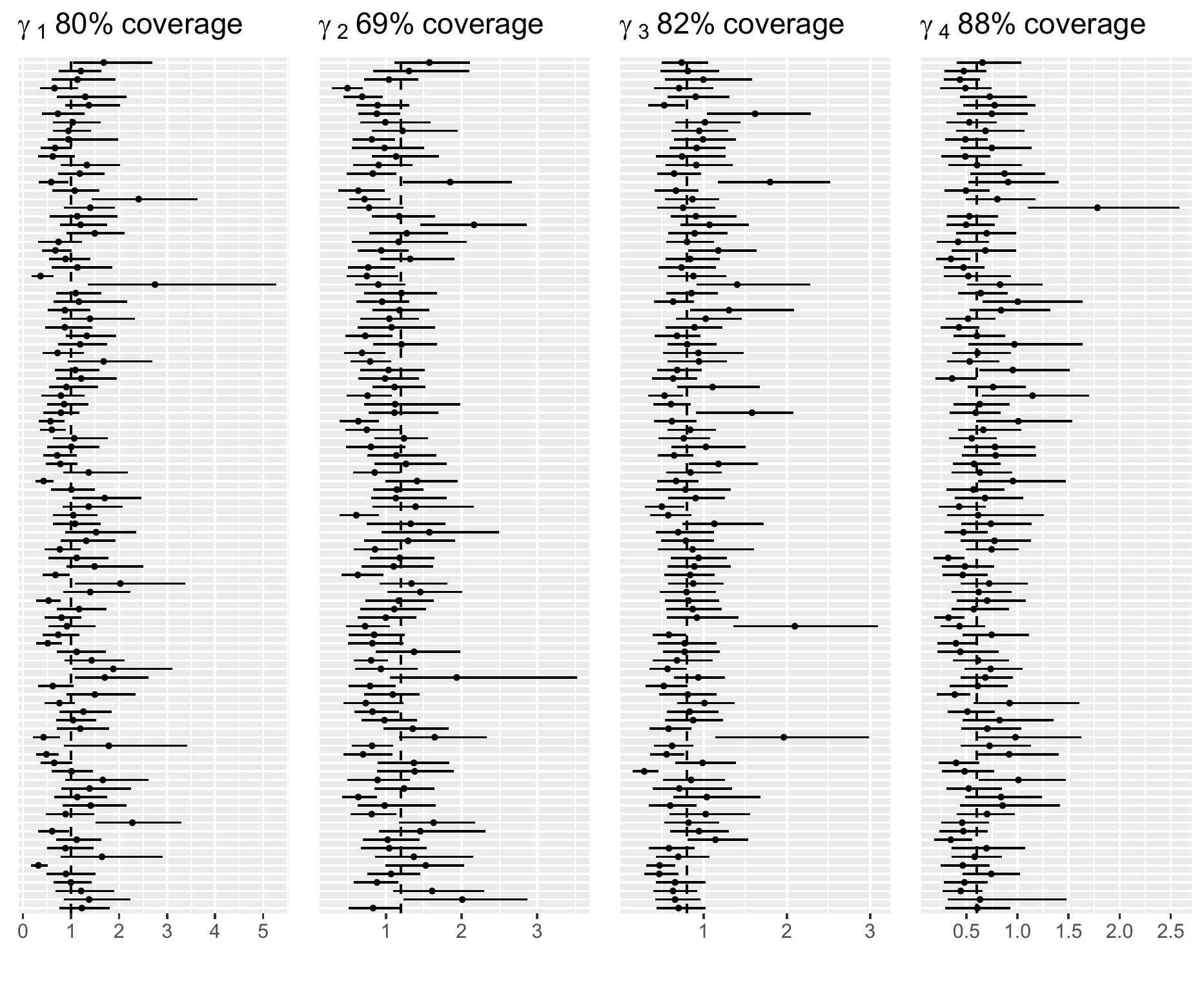}
	\includegraphics[width=.7\linewidth]{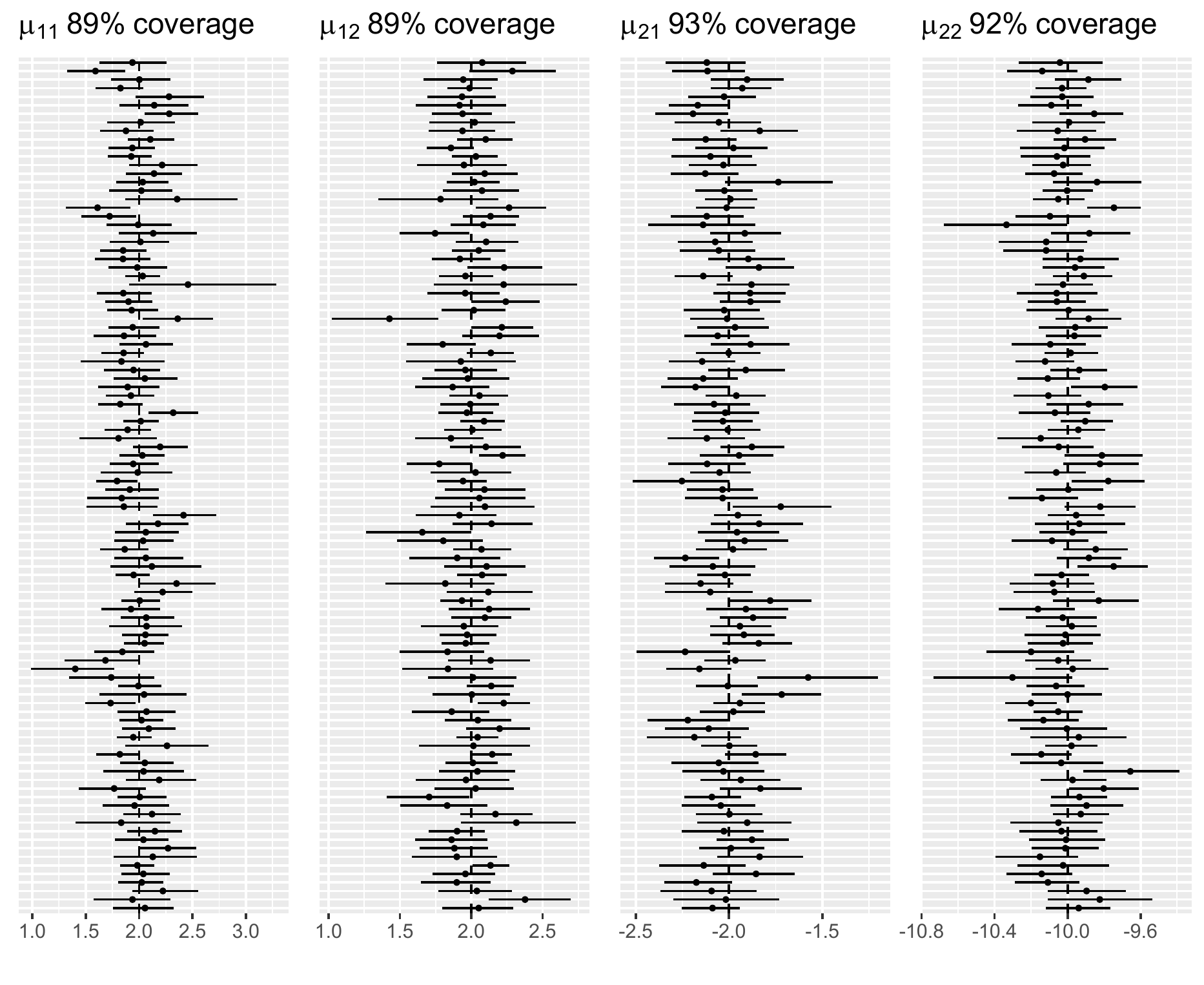}
\end{figure*}
\begin{figure*}[h]
	\centering
	\includegraphics[width=.8\linewidth]{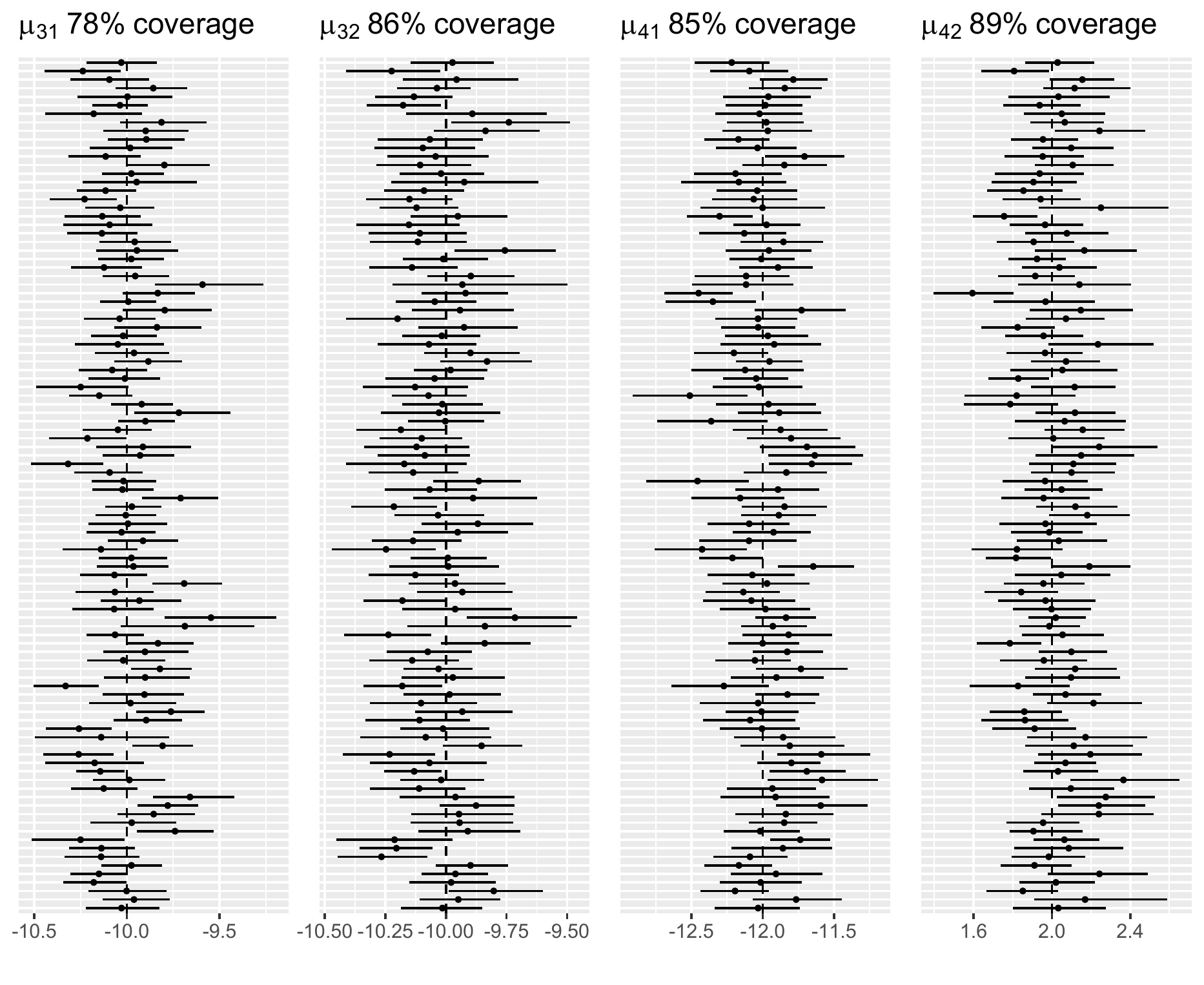}
	\includegraphics[width=.8\linewidth]{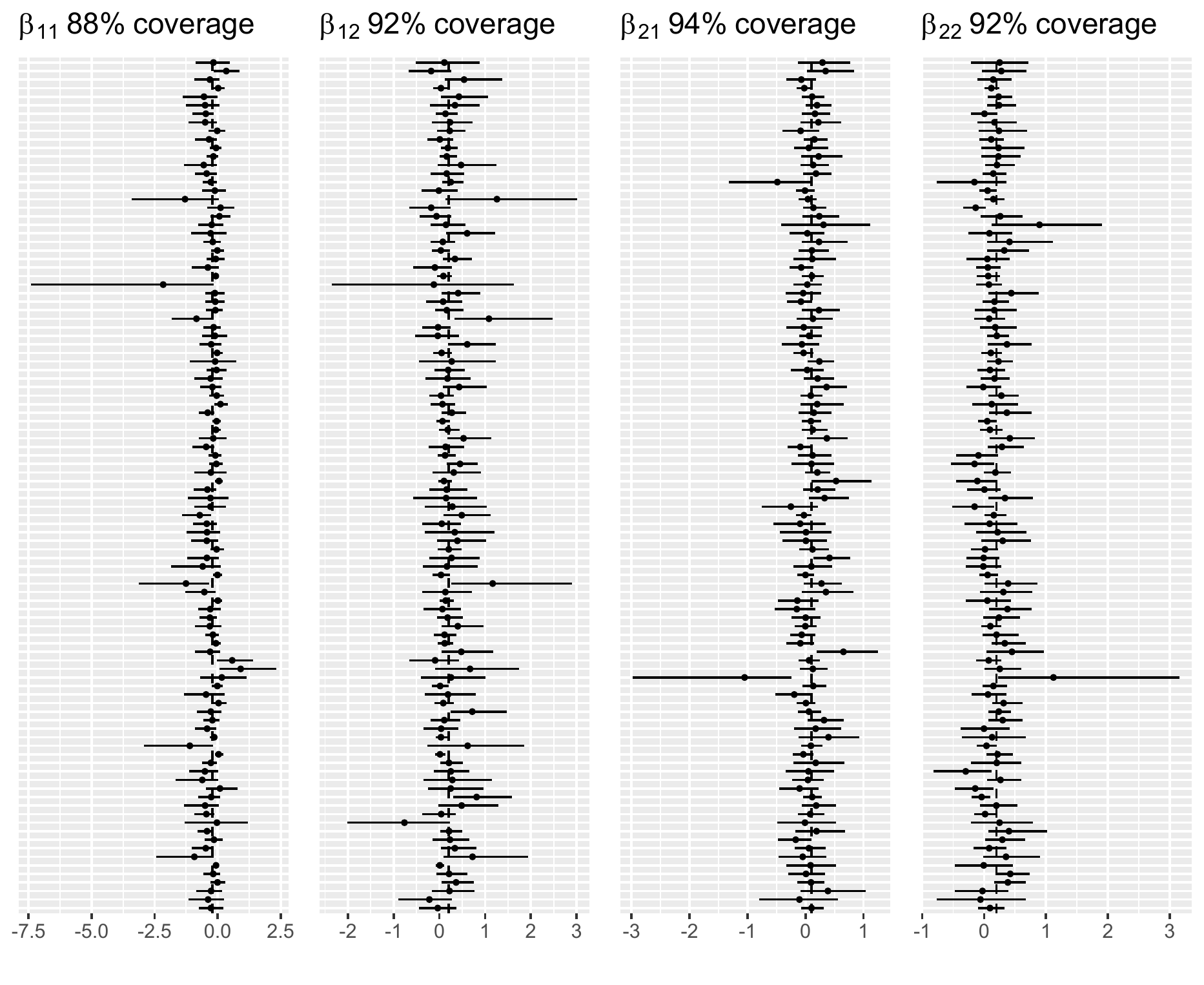}
\end{figure*}
\begin{figure*}[h]
	\centering
	\includegraphics[width=.8\linewidth]{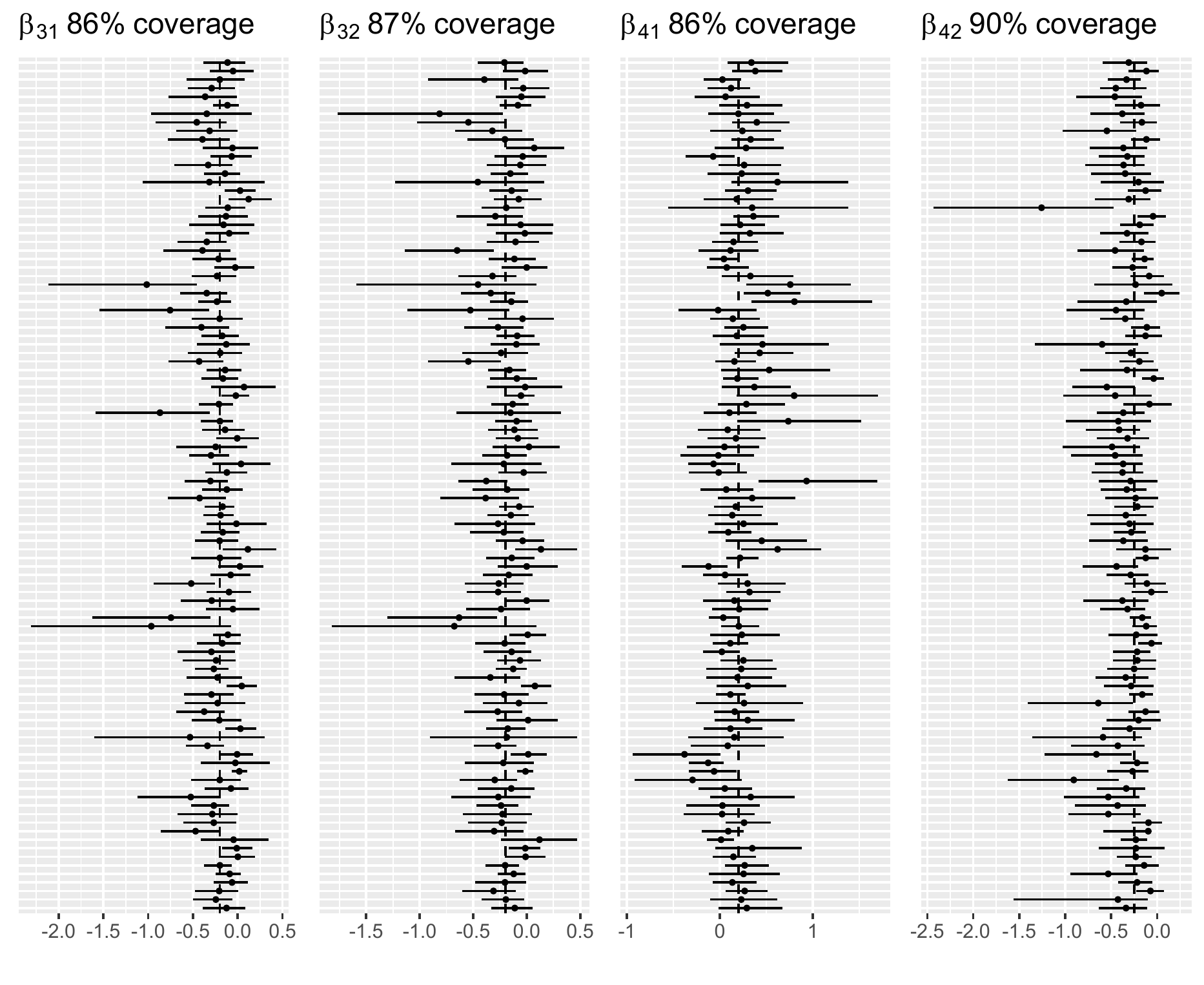}
	\includegraphics[width=.8\linewidth]{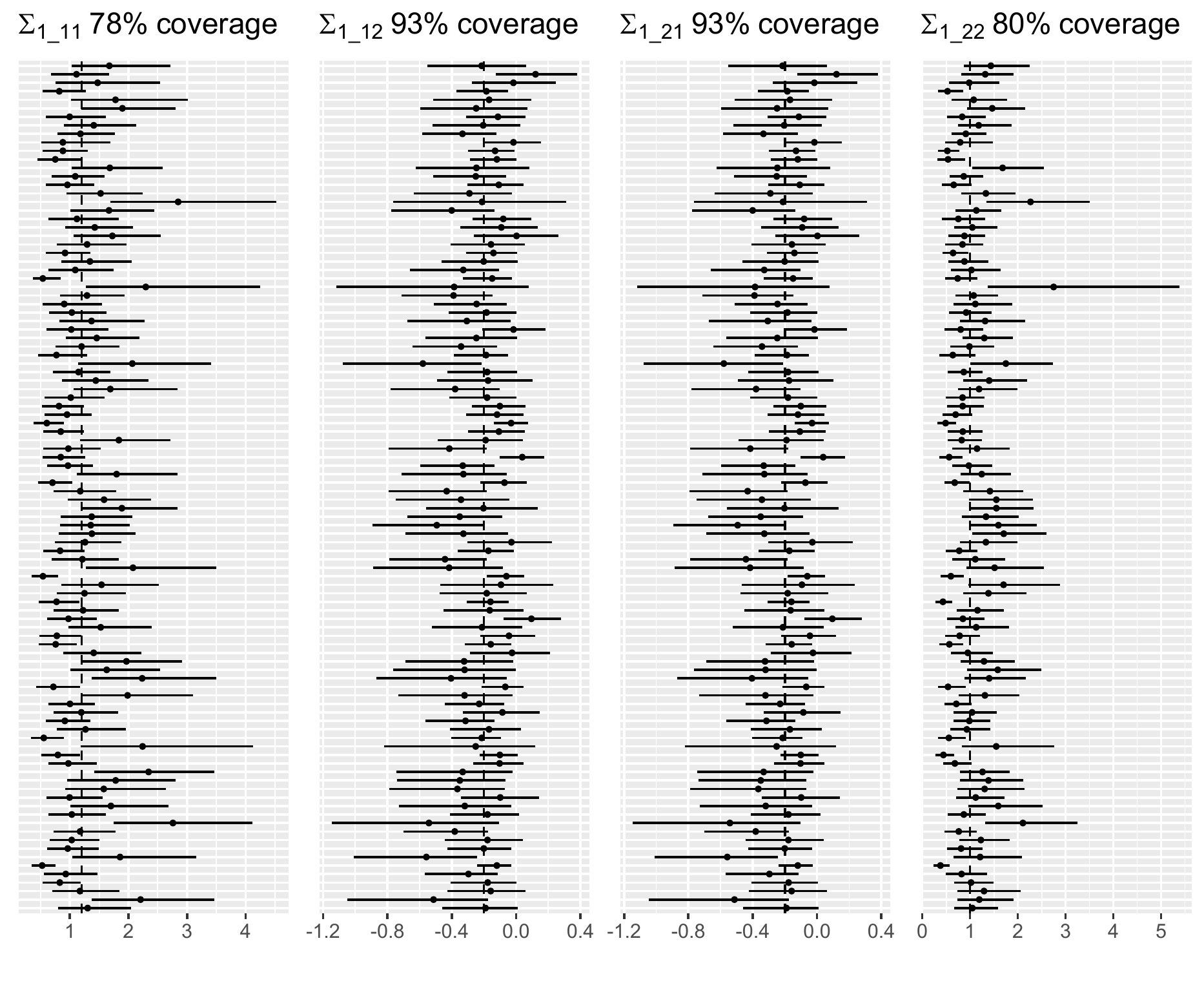}
\end{figure*}
\begin{figure*}[h]
	\centering
	\includegraphics[width=.8\linewidth]{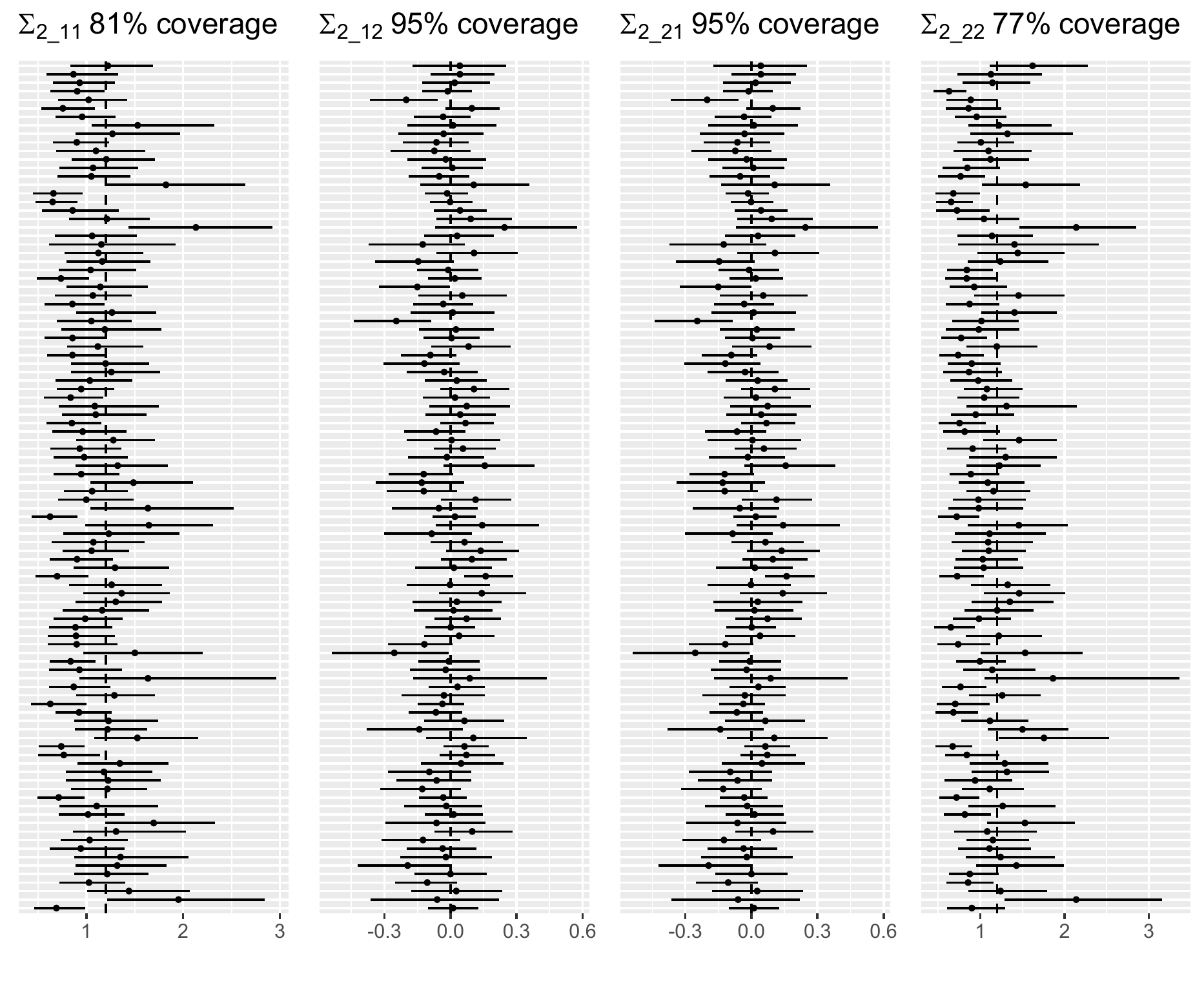}
	\includegraphics[width=.8\linewidth]{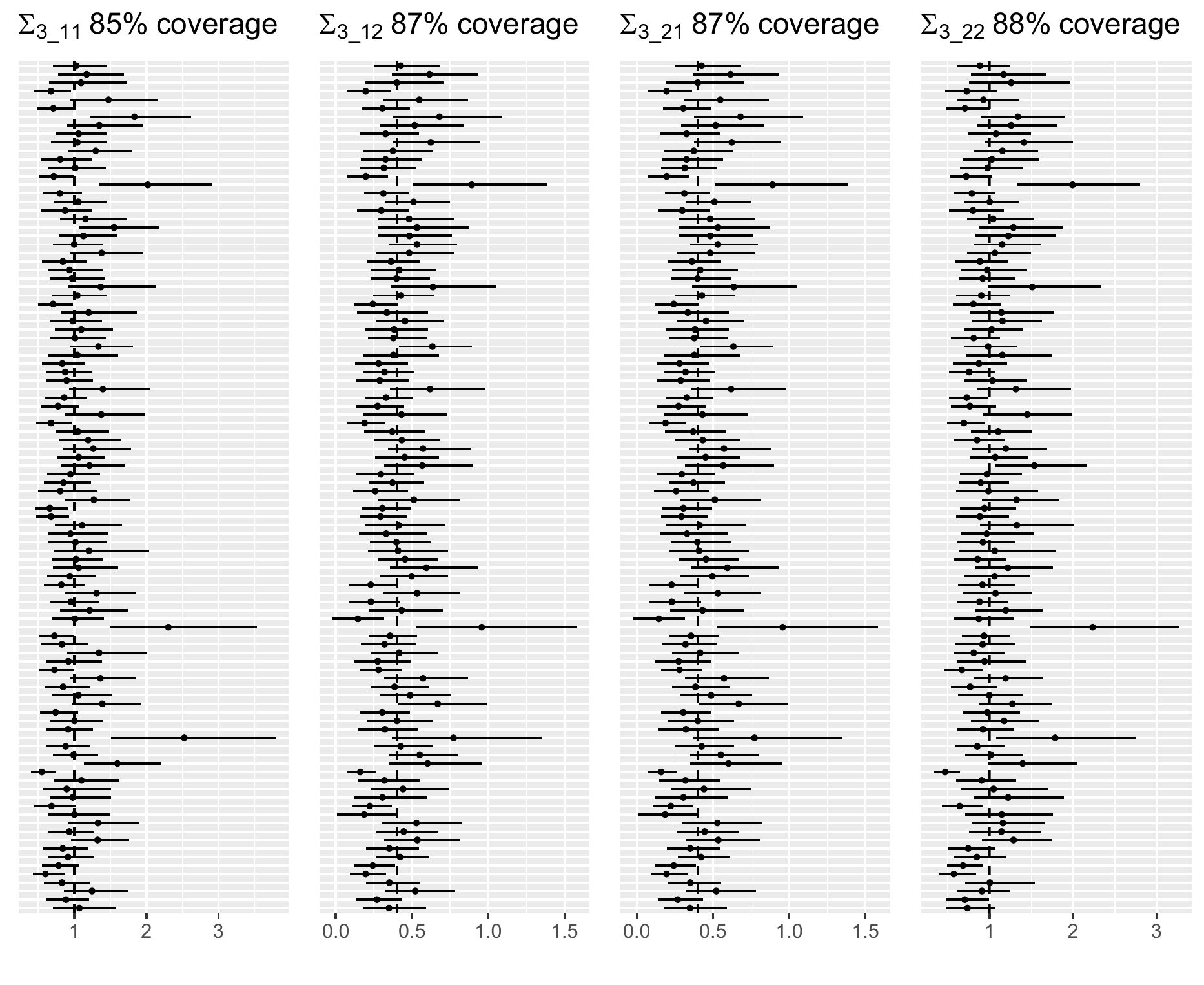}
\end{figure*}
\begin{figure*}[h]
	\centering
	\includegraphics[width=.8\linewidth]{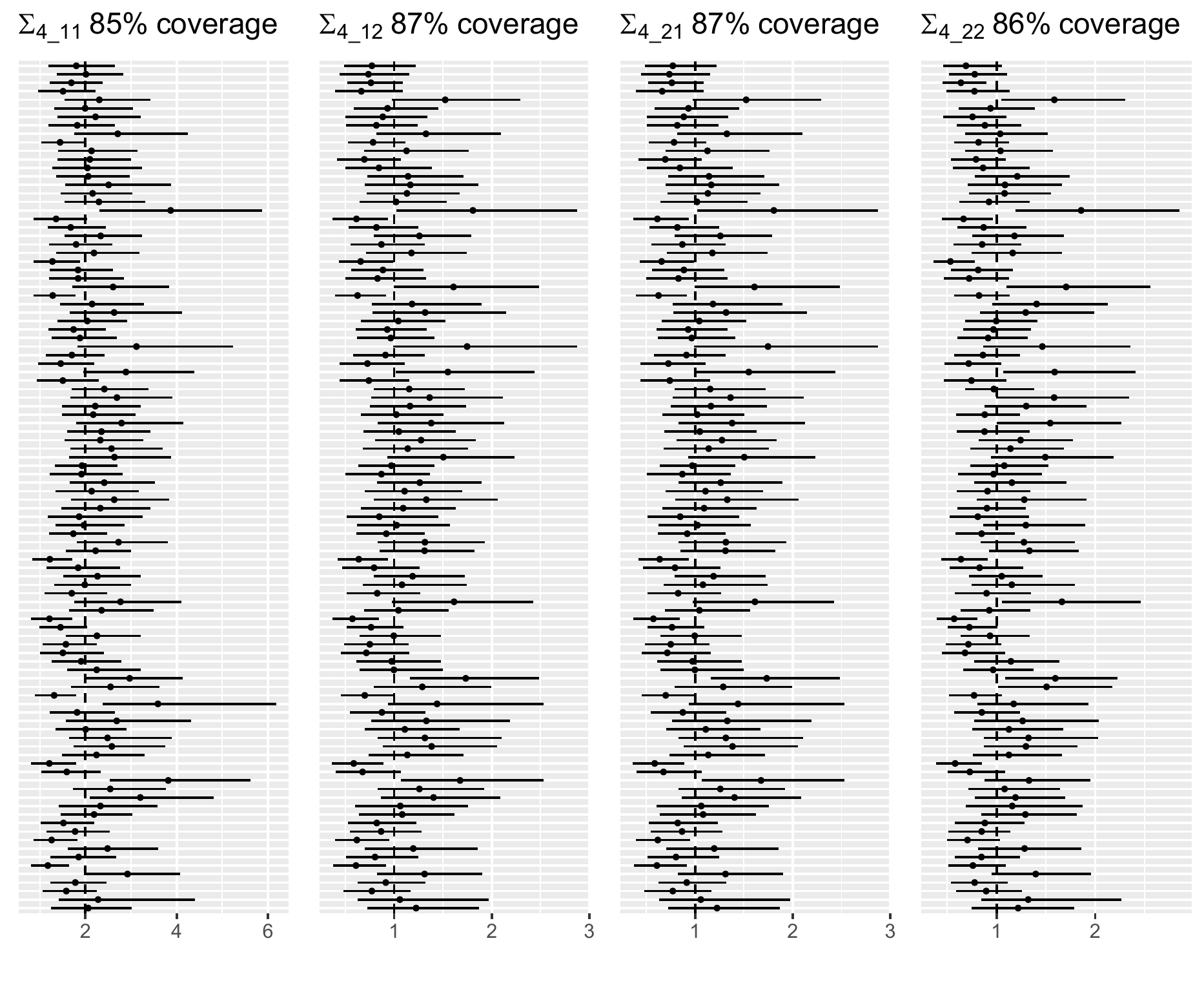}
	\includegraphics[width=.8\linewidth]{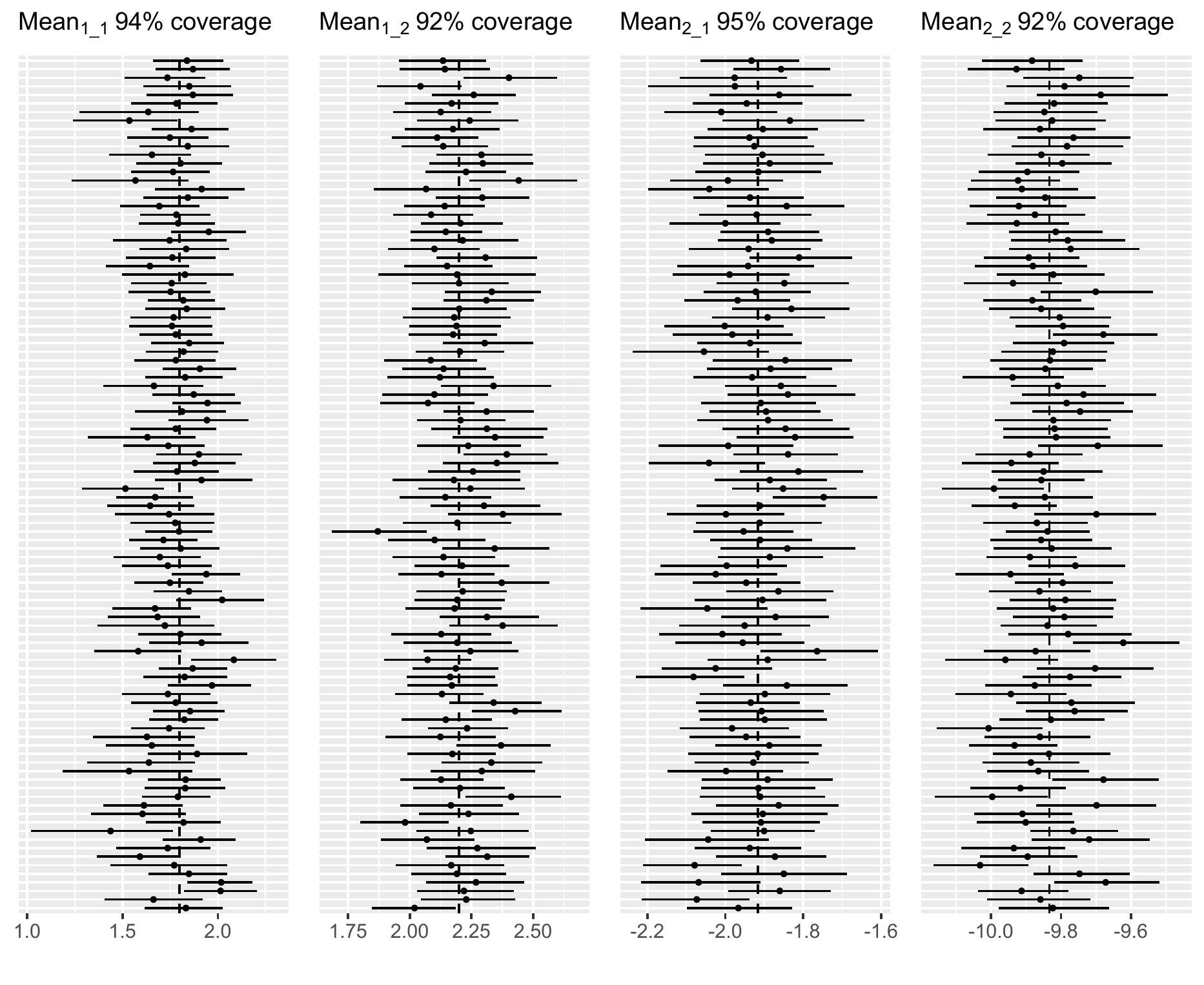}
\end{figure*}
\begin{figure*}[h]
	\centering
	\includegraphics[width=.8\linewidth]{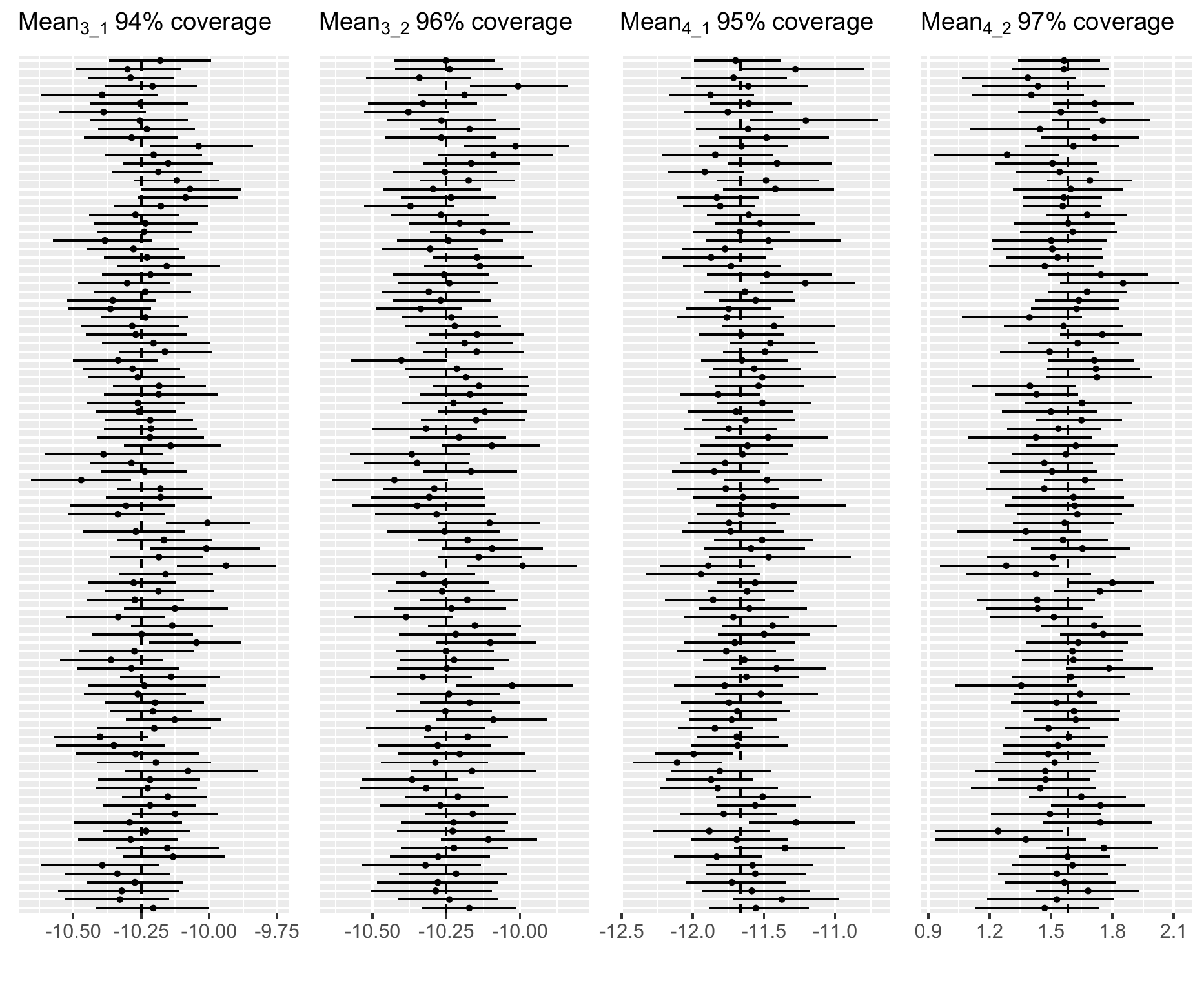}
	\includegraphics[width=.8\linewidth]{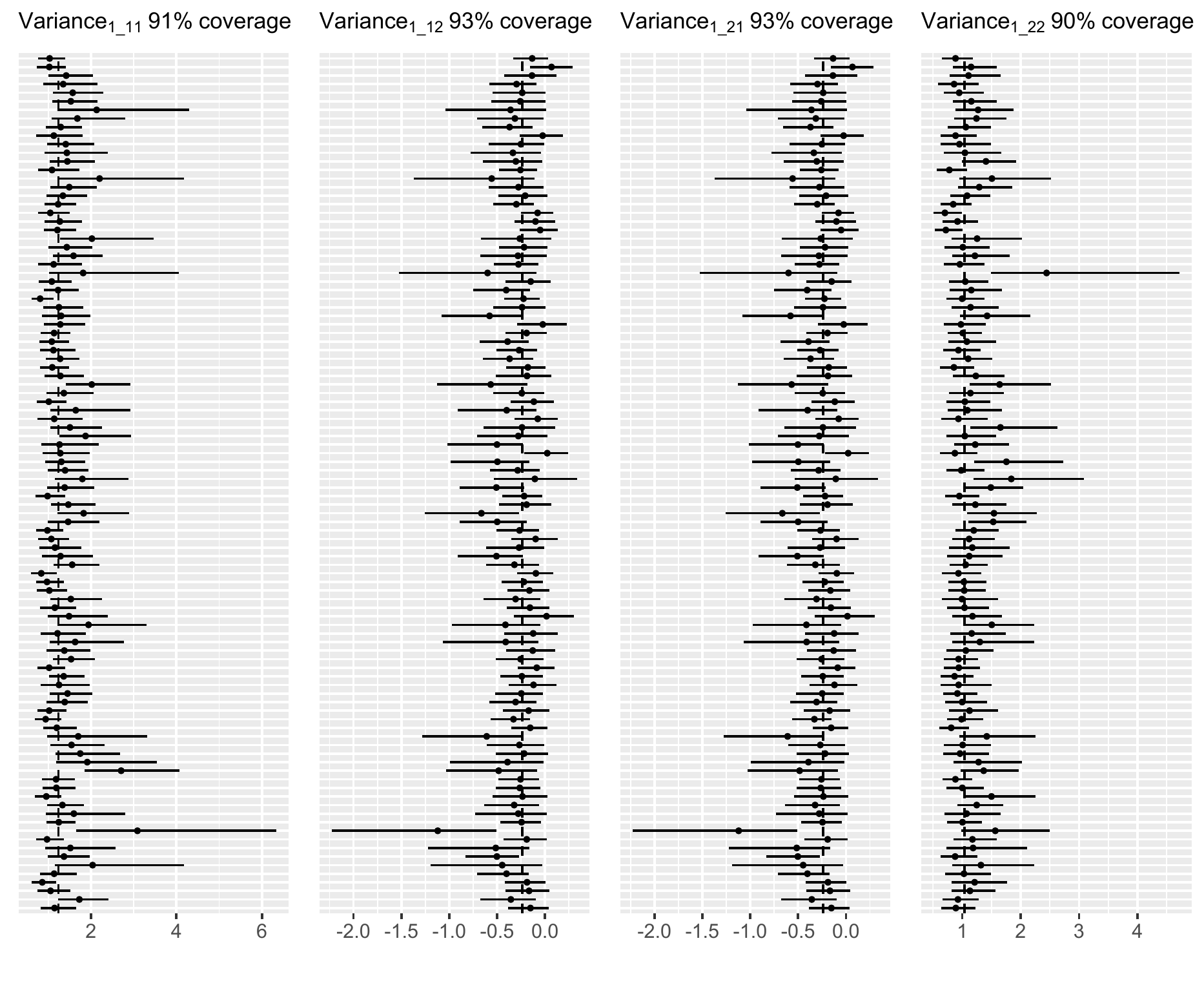}
\end{figure*}
\begin{figure*}[h]
	\centering
	\includegraphics[width=.8\linewidth]{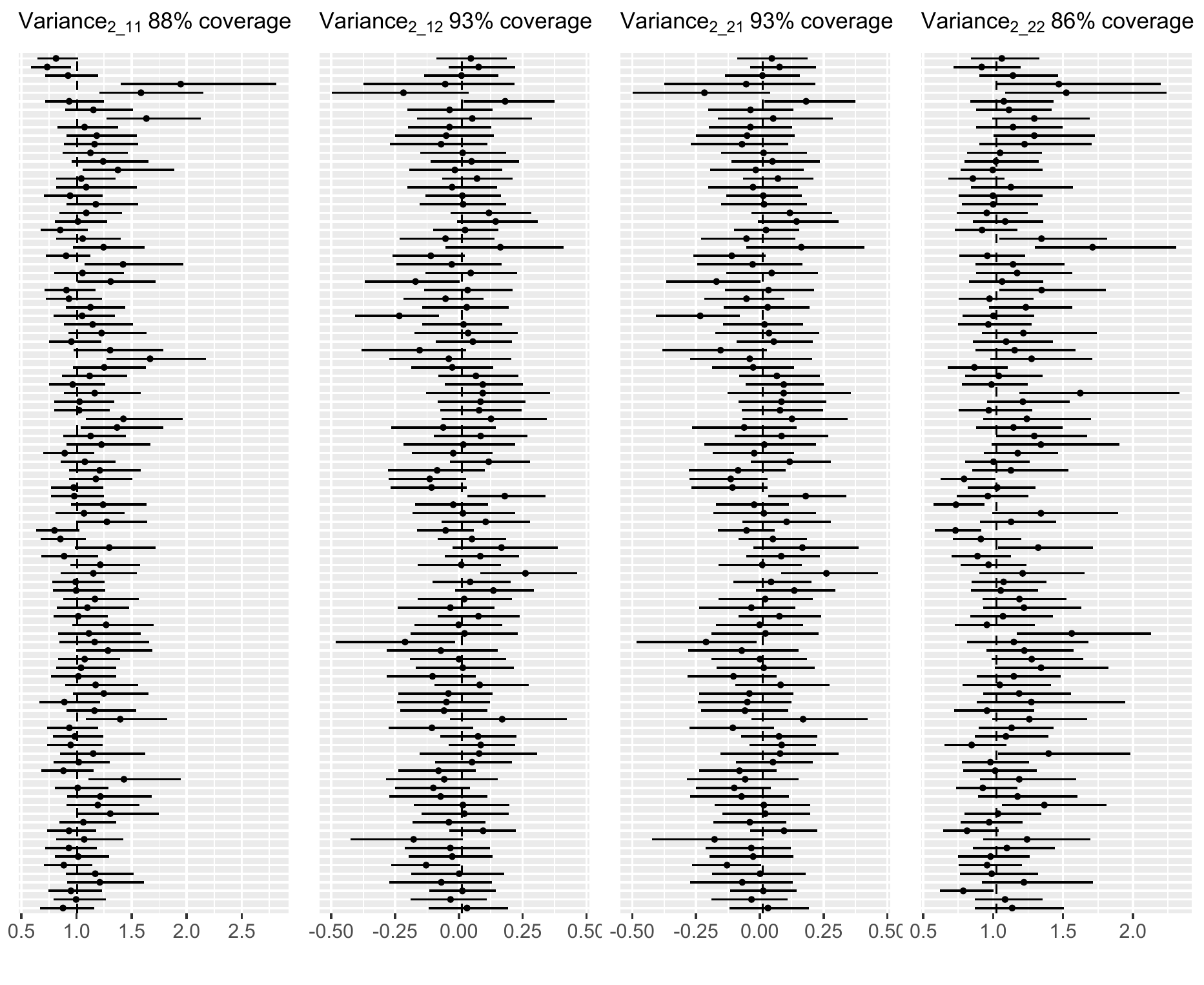}
	\includegraphics[width=.8\linewidth]{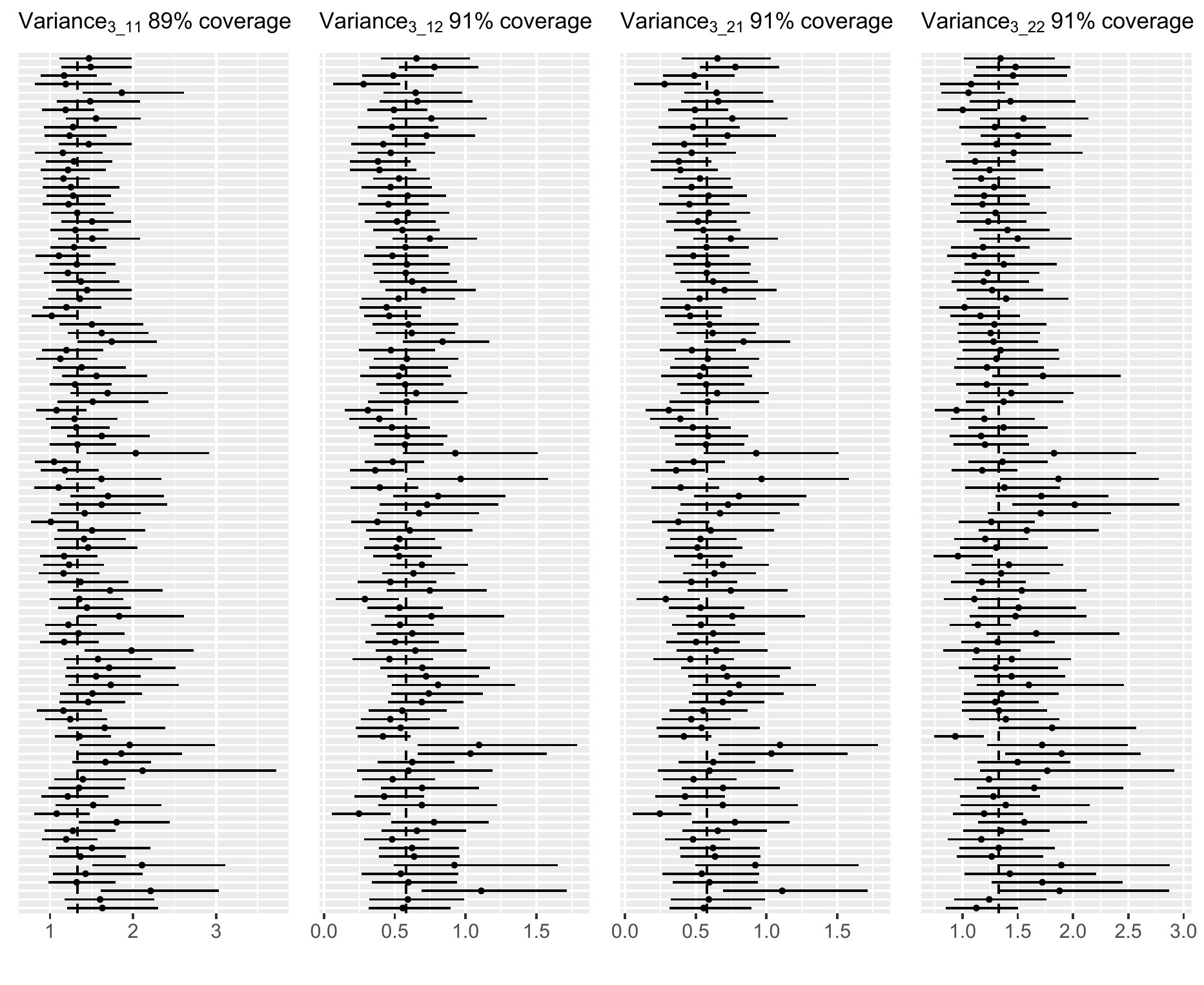}
\end{figure*}
\begin{figure*}[h]
	\centering
	\includegraphics[width=.8\linewidth]{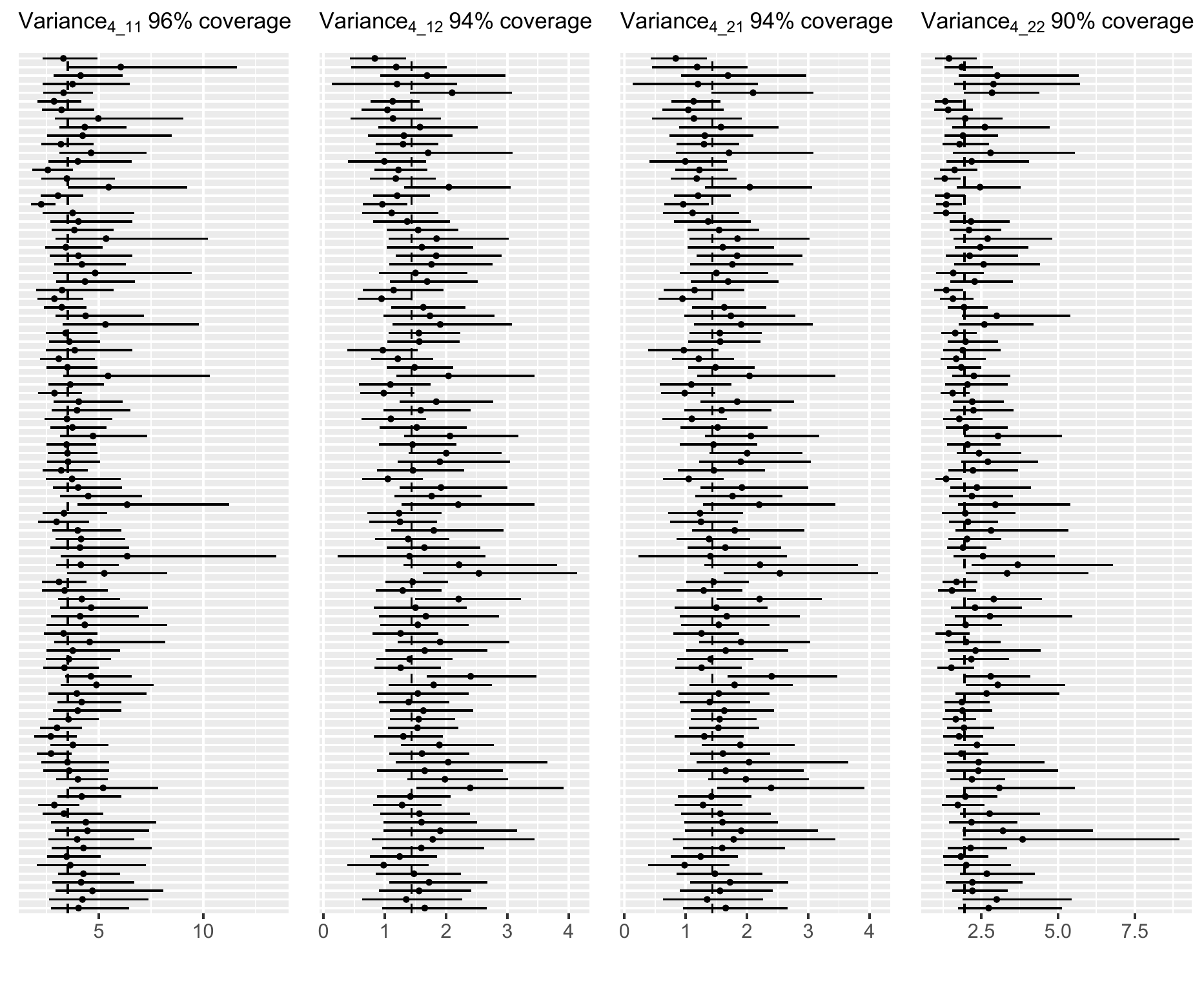}
	\caption{95\% Credible Intervals for estimated parameters for all 100 runs for Simulation Study 1.}
	\label{fig:parameter_ci}
\end{figure*}

\end{document}